\documentstyle[12pt]{article}
\begin{document}
\vspace*{-1.5cm}
\begin{flushright}
{\it Preprint}~ IC/94/352
\end{flushright}
\begin{center}
{\large {\bf FINITE--DIMENSIONAL REPRESENTATIONS OF\\
THE QUANTUM SUPERALGEBRA U$_{q}$[gl(2/2)] II:\\
\vspace*{2mm}
NONTYPICAL REPRESENTATIONS AT GENERIC}}
{}~{\Large{\bf $q$}}\\
\vskip 0.8truecm
\def\thefootnote{a)} {\bf Nguyen Anh Ky}
\footnote{~Fellow
of the Ministry of Education and Training, Hanoi, {\bf Vietnam}}
\def\thefootnote{ }
\footnote{~Permanent Mailing Address:
Centre for Polytechnic Educations,
Vinh, {\bf Vietnam}}\\
\normalsize International Centre for Theoretical Physics\\
P.O. Box 586,
Trieste 34100, {\bf Italy}\\
\vskip 0.4truecm
\normalsize and
\vskip 0.4truecm
{\bf Nedjalka I. Stoilova}\\
\normalsize Institute for Nuclear Research and Nuclear Energy\\
Blvd.
Tsarigradsko chaussee 72, Sofia 1784, {\bf Bulgaria}
\vskip 1truecm
\end{center}
\begin{center}
{\bf Abstract} \\[1mm]
\end{center}

    The construction approach proposed in the previous paper
Ref. 1 allows us there and in the present paper to construct
at generic deformation parameter $q$ all finite--dimensional
representations of the quantum Lie superalgebra $U_{q}[gl(2/2)]$.
The finite--dimensional $U_{q}[gl(2/2)]$-modules $W^{q}$
constructed in Ref. 1 are either irreducible or indecomposible.
If a module $W^{q}$ is indecomposible, i.e. when the condition
(4.41) in Ref. 1 does not hold, there exists an invariant maximal
submodule of $W^{q}$, to say $I_{k}^{q}$, such that the
factor-representation in the factor-module $W^{q}/I_{k}^{q}$ is
irreducible and called nontypical. Here, in this paper,
indecomposible representations and nontypical finite--dimensional
representations of the quantum Lie superalgebra $U_{q}[gl(2/2)]$
are considered and classified as their module structures are
analized and the matrix elements of all nontypical representations
are written down explicitly.
\\[1cm]
\newpage
\begin{flushleft}
{\bf I. INTRODUCTION}
\end{flushleft}
\vspace*{2mm}

     As mentioned in Ref. 1 (referred hereafter to as
{\bf I}), explicit representations
(at generic $q$) are known only for some quantum
superalgebras of particular types like
$U_{q}[osp(1/2)]$, $^{2}$  $U_{q}[gl(1/n)]$, $^{3}$, etc.,
while for higher rank quantum superalgebras
of nonparticular type, besides some
q-oscillator representations (see, for example, Refs. 4),
other representations had not been explicitly constructed
and completely investigated (at neither generic $q$ nor
$q$ being roots of unity) with the exception of the
module structure (and some general aspects) $^{5}$
and a class of representations $^{6}$
of the quantum superalgebra $U_{q}[gl(m/n)]$.
In this paper we continue our investigations on
finite--dimensional representations of the quantum Lie
superalgebra $U_{q}[gl(2/2)]$ $U_{q}[gl(2/2)]$ started
in {\bf I}
where a procedure for their explicit construction
was proposed.
Especially, all the typical representations of
$U_{q}[gl(2/2)]$ were studied completely and constructed
explicitly.
This paper is devoted to indecomposible
representations and nontypical representations of
$U_{q}[gl(2/2)]$ which are classified in 5 classes.
Nontypical representations of every class are investigated
detailly as the matrix elements are presented in their
explicit forms. In such a way, in {\bf I} and the present
paper, $U_{q}[gl(2/2)]$ becomes, to our best knowledge,
the highest rank superalgebra
of a nonparticular type $U_{q}[gl(m/n)]$, $m,n \geq 2$,
with all finite--dimensional representations constructed
explicitly at generic $q$, .\\

   Introduced by Drinfel'd $^{7}$ and Jimbo $^{8}$ the
quantum deformation (q-deformation) of universal enveloping
(super) algebras is one of four approaches to defining
quantum (super) groups $^{9-11}$. In particular, the quantum
Lie superalgebra $U_{q}[gl(2/2)]$ as a quantum deformation
of the universal enveloping algebra $U[gl(2/2)]$ is
completely generated by the Cartan--Chevalley generators
obeying a number of defining relations (see {\bf I} and
Sect. II), namely,
the Cartan--Kac supercommutation relations, the Serre
relations and the extra--Serre relations $^{12}$.
Representing a quantum extension of the induced
representation method developed by Kac $^{13}$ in the case
of Lie superalgebras (from now on, only superalgebras) the
approach proposed in {\bf I} allows us to construct at
generic deformation parameter $q$ all finite--dimensional
representations of the quantum superalgebra $U_{q}[gl(2/2)]$
(and, certainly, is applicable to other higher rank quantum
(super) groups).
The induced $U_{q}[gl(2/2)]$--modules $W^{q}$ constructed in
{\bf I} are either irreducible or indecomposible (see {\bf I},
Proposition 2). Every
(typical and nontypical) $U_{q}[gl(2/2)]$--module $W^{q}$
is decomposed into a direct sum of 16 or less irreducible
$U_{q}[gl(2)\oplus gl(2)]$-modules $V^{q}_{k}$,
$0\leq k\leq 15$,
$$W^{q}=
\bigoplus_{k=0}^{15}V_{k}^{q}.\eqno(1.1)$$
Moreover, the module structure (1.1) of $W^{q}$
reminds us of the module structure of the classical
$gl(2/2)$--modules $W$ constructed in Ref. 14 (see
also Ref. 15 especially for nontypical modules $W$)
and decomposed into a direct sum of 16, at most,
finite--dimensional irreducible $gl(2) \oplus gl(2)$.
For a basis of $W^{q}$ we can choose, as explained
in {\bf I}, a union of the Gel'fand--Zetlin (GZ) basises
of all the submodules $V^{q}_{i}$ included in the
decomposition (1.1) of $W^{q}$. In {\bf I} all the
analyses and the matrix elements were provided in
the framework of such a basis which we suggest to be called
the quasi--Gel'fand--Zetlin (QGZ) basis. The latter,
as emphasized also in Ref. 1,
is convenient for a construction of finite--dimensional
(typical and nontypical) $U_{q}[gl(2/2)]$-modules and
spanned on all possible patterns of the form
$$\left[
\begin{array}{lccc}

                  m_{13}& m_{23}&  m_{33}& m_{43} \\

                  m_{12}& m_{22}&  m_{32}&  m_{42}\\
                  m_{11}&   0   &  m_{31}&    0
\end{array}
\right],\eqno(1.2)$$
where  $m_{i,j}$ are complex numbers such that
{}~$m_{13} - m_{23}$,~  $m_{33}- m_{43}$,~
$m_{12}-m_{11}$,~ $m_{11}-m_{22}$,~ $m_{32}-m_{31}$
{}~and~  $m_{31}-m_{42}$~ are non-negative integers:
$$m_{13},~ m_{23},~ m_{33},~ m_{43} \in {\bf C},$$
$$m_{13} - m_{23},~  m_{33}- m_{43},~
m_{12}-m_{11}, ~m_{11}-m_{22},~ m_{32}-m_{31},~
m_{31}-m_{42} \in {\bf Z}_{+}. \eqno(1.3)$$
The patterns (1.2) for a fixed second row
$$[m]_{k}:= [m_{12},m_{22},m_{32},m_{42}]_{k}
\eqno(1.4)$$
describe
a GZ basis of a finite--dimensional irreducible module:
$$V^{q}_{k}\equiv V^{q}_{i}([m]_{k})\eqno(1.5)$$
of the quantum subalgebra
$U_{q}[gl(2)\oplus gl(2)]$, while the first  row
$$[m]\equiv [m_{13},m_{23},m_{33},m_{43}]\eqno(1.6)$$
does not change through out the module $W^{q}$ and
characterizes the latter as the whole:
$$W^{q}\equiv W^{q}([m])=
\bigoplus_{k=0}^{15}V_{k}^{q}([m]_{k}).\eqno(1.7)$$
The basis vectors (1.2) and the subspaces $V^{q}_{k}$
will be specified explicitly later in Sec. II.
Thus, all the even Cartan--Chevalley generators
of $U_{q}[gl(2/2)]$ can shift, at most, only the third
row $[m_{11},0,m_{31},0]$, while the odd
generators of $U_{q}[gl(2/2)]$ leave always invariant
only the first row $[m]$. In other words, we can say
that the  basis description (1.2) relaxing the original
Gel'fand--Zetlin basis description $^{16}$ corresponds
to the branching rule
$$U_{q}[gl(2/2)]\supset U_{q}[gl(2)\oplus gl(2)]\supset
U_{q}[gl(1)\oplus gl(1)]\eqno(1.8)$$
This is the reason of why the basis (1.2) is referred
to as a quasi - Gel'fand - Zetlin basis. The basis
description (1.2) was introduced for the case of the
classical superalgebra  $gl(2/2)$ in Ref. 14
(see also Ref. 15) and extended in {\bf I} to
the case of the quantum superalgebra $U_{q}[gl(2/2)]$.
For some other extensions of the GZ basis description
will be discussed in the Conclusion.\\

   An induced module $W^{q}$ and, therefore,
the corresponding representation that $W^{q}$ carries
are irreducible and called typical if and only if the
condition (I.4.41) (formula (4.41) in {\bf I})
$$(l_{13}+l_{33}+3)(l_{13}+l_{43}+3)(l_{23}+l_{33}+3)
(l_{23}+l_{43}+3)\neq 0\eqno(1.9)$$
holds, where $l_{ij}=m_{ij}-i$, for $i=1,2$ and
$l_{ij}=m_{ij}-i+2$, for  $i=3,4$.
The typical modules and representations of
$U_{q}[gl(2/2)]$ were detailly investigated and all
explicitly constructed at generic deformation parameter
$q$ in {\bf I}. When the above-mentioned condition (1.9)
(or I.4.41) is violated the modules $W^{q}$ are no
longer irreducible but indecomposible. It turns out that
the indecomposible representations are divided into
five classes $k=1,2,3,4,5$ (see (3.1)-(3.5)). Then in
each indecomposible module $W^{q}$, belonging always to
one of these five classes $k$, there exists a maximal
invariant submodule, to say $I^{q}_{k}$, and the
compliment to $I^{q}_{k}$ subspace of $W^{q}$ is not
invariant under $U_{q}[gl(2/2)]$ transformations.
The factor module $W^{q}_{k}=W^{q}/I^{q}_{k}$, however,
carries an irreducible representation of $U_{q}[gl(2/2)]$
which is called nontypical (cf. Ref. 15). Here,
following a programme close to the classical one $^{14,15}$
we shall determine all nontypical $U_{q}[gl(2/2)]$-modules
and choose within every of them an appropriate basis so
that the decompositions of the nontypical representations
of $U_{q}[gl(2/2)]$ into irreducible representations of
$U_{q}[gl(2)\oplus gl(2)]$ are evident and their explicit
matrix elements can more easily be written down. Going step
by step we can imagine that the present paper is something
like a quantum deformation of Ref. 15, referred hereafter
to as {\bf I$^{\ast}$}. (Therefore, our plan
here, in general, will go hand in hand with that of
{\bf I$^{\ast}$}.)\\

    Since for an explicit construction of the
finite--dimensional representations of $U_{q}[gl(2/2)]$
we have already made in {\bf I} a relevant introduction
to the problem, here, in Section II, in order to make
the present paper more self-contained, we repeat only
briefly some of the basic concepts and the main points
in defining the quantum superalgebra $U_{q}[gl(2/2)]$
and its induced representations. According to the plan,
we shall consider indecomposible representations of
$U_{q}[gl(2/2)]$ in Section III where all the nontypical
representations are classified and constructed explicitly
at generic $q$. In Section IV we
shall show that the class of the modules $W^{q}$
determined in {\bf I} and in the present paper, contains
all finite--dimensional irreducible modules of
$U_{q}[gl(2/2)]$.
The conclusion is made in Section V where we state that
the finite--dimensional irreducible representations
of $U_{q}[gl(2/2)]$ are quantum deformations of the
finite--dimensional irreducible representations
of the classical $gl(2/2)$. In the Appendix we write
down the matrix elements of the generators $E_{23}$
and $E_{32}$ for the nontypical representations of all
the classes with the exception of the class 1 which is
described more detailly in Sec. III.\\

  Through out the paper, for a convenient reading we shall
keep as many as possible of the abbreviations and notations
used in Ref. 1,14 and 15 among the following ones:
\begin{tabbing}
\=12345678\=$V_{l}\otimes V_{r}$ -- \=tensor product
between two linear spaces
$V_{l}$ and $V_{r}$\= or a tensor product \=\kill
\>\>fidirmod(s) -- finite--dimensional
irreducible module(s), \\[2mm]
\>\>GZ basis -- Gel'fand--Zetlin basis,\\[2mm]
\>\>QGZ basis -- quasi-- Gel'fand--Zetlin basis,\\[2mm]
\>\>lin.env.\{X\} -- linear envelope of X,\\[2mm]
\>\>$q$ -- the deformation parameter,\\[2mm]
\>\>$V^{q}_{l}\otimes V^{q}_{r}$ -- tensor product
between two linear spaces
$V^{q}_{l}$ and $V^{q}_{r}$\\
\>\>\>~~or a tensor product between a
$U_{q}[gl(2)_{l}]$-module $V^{q}_{l}$\\
\>\>\>~~and a $U_{q}[gl(2)_{r}]$-module
$V^{q}_{r}$,\\[2mm]
\>\>$T^{q}\odot V^{q}_{0}$ -- tensor
product between two $U_{q}[gl(2)\oplus
gl(2)]$-modules\\
\>\>\>~~$T^{q}$ and $V^{q}_{0}$,\\[2mm]
\>\>$[x]_{q}={q^{x}-q^{-x}\over q-q^{-1}}$,
{}~ where $x$ is some
number or operator,\\[2mm]
\>\>$[x]\equiv [x]_{q^{2}}$,\\[2mm]
\>\>$[E,F\}$ -- supercommutator between $E$
and $F$,\\[2mm]
\>\>$[E,F\}_{q}\equiv EF-qFE$ -- q-deformed
supercommutator between $E$
and $F$,\\[2mm]
\>\>$[m]=[m_{13},m_{23},m_{33},m_{43}]$ -- the highest weight,\\[2mm]
\>\>$l_{ij}=m_{ij}-i$ for $i=1,2$ and
 $l_{ij}=m_{ij}-i+2$ for $i=3,4$,\\[2mm]
\>\>$I_{k}^{q}$ -- the maximal invariant subspace in $W^{q}([m])$,
corresponding to the class $k$,\\[2mm]
\>\>~~~~~ $k=1,2,3,4,5$
(see (3.1)-(3.5)),\\[2mm]
\>\>$W^{q}_{k}([m])=W^{q}([m])/I_{k}^{q}$ --
the class $k$ nontypical module,\\[2mm]
\>\>$(m;~ m_{kl} = \alpha)^{\pm ij}$ -- a pattern obtained
from $(m)$ by replacing $m_{kl}$ with $\alpha$\\[2mm]
\>\>\>~~~~~~~~~~~ and by shifting
$m_{ij} \rightarrow m_{ij} \pm 1$,\\[2mm]
\>\>$[m;~ m_{kl} = \alpha]$ -- a signature obtained
from $[m]$ by replacing $m_{kl}$ with $\alpha$,\\[2mm]
\>\>${\frame{\shortstack{{\small ~-a,-b,c,d}}~}}_{~(s)}$ --
a ~$U_{q}[gl(2/2)]$-fidirmod with a signature\\[2mm]
\>\>\>~~~~~
$[m_{13}-a,m_{23}-b,m_{33}+c,m_{43}+d]$
{}~(see (3.7)),\\[2mm]
\>\>$\stackrel{{\small inv}}
{\frame{\shortstack{
{\small ~-a,-b,c,d}}~}}_{~(s)}$ --
a ~${\frame{\shortstack{{\small ~-a,-b,c,d}}~}}_{~(s)}$
%%a $U_{q}[gl(2/2)]$-fidirmod with a signature\\[2mm]
%%\>\>\>$[m_{13}-a,m_{23}-b,m_{33}+c,m_{43}+d]$
belonging
to $I_{k}^{q}$ (see (3.8)),\\[2mm]
\>\>$\stackrel{{\small noninv}}
{\frame{\shortstack{
{\small ~-a,-b,c,d}}~}}_{~(s)}$ --
a ~${\frame{\shortstack{{\small ~-a,-b,c,d}}~}}_{~(s)}$
%%a $U_{q}[gl(2/2)]$-fidirmod with a signature\\[2mm]
%%\>\>\> $[m_{13}-a,m_{23}-b,m_{33}+c,m_{43}+d]$
belonging to the compliments\\[2mm]
\>\>\>~~~~~
to $I_{k}^{q}$ subspaces in $W^{q}$
(see (3.9)).
\end{tabbing}
 Note that the quantum deformation
$[x]\equiv [x]_{q^{2}}$ of $x$ must not be confused with
the highest weight (signature) $[m]$ in the (quasi) GZ
basis $(m)$ or with the notation $[ ~,~ ]$ for commutators.
\\[5mm]
\begin{flushleft}
{\bf II. U$_{q}$[gl(2/2)] AND ITS
INDUCED REPRESENTATIONS $^{1}$}
\end{flushleft}
\vspace*{2mm}

  The quantum superalgebra $U_{q}[gl(2/2)]$ $^{1}$
as a quantum deformation of the universal
enveloping algebra $U[gl(2/2)]$ of the superalgebra
$gl(2/2)$ is completely generated by
the Cartan--Chevalley generators
$E_{ii}$, $i=1,2,3,4$, $E_{12}\equiv
e_{1}$, $E_{23}\equiv e_{2}$,
$E_{34}\equiv e_{3}$, $E_{21}\equiv
f_{1}$, $E_{32}\equiv f_{2}$ and
$E_{43}\equiv f_{3}$ satisfying  $^{1}$
\begin{tabbing}
\=1234567891234567891\=$[E_{ii},E_{jj}]$12345\= =1
\= 0,1234
\=$[E_{ii},E_{j,j+1}]$\= =
\=$(\delta_{ij}-\delta_{i,j+1})E_{j,j+1}$1\=\kill

{}~~~~a) the Cartan-Kac supercommutation relations
($1\leq i,i+1,j,j+1\leq 4$):\\[2mm]
\>\>$[E_{ii},E_{jj}]$\> = \>0,\\[1mm]
\>\>$[E_{ii},E_{j,j+1}]$\>=\>
$(\delta_{ij}-\delta_{i,j+1})E_{j,j+1}$,\\[1mm]
\>\>$[E_{ii},E_{j+1,j}]$\>=\>
$(\delta_{i,j+1}-\delta_{ij})E_{j+1,j}$,\\[1mm]
\>\>$[E_{i,i+1},E_{j+1,j}\}$\>=\>
$\delta_{ij}[h_{i}]_{q^{2}}$,
{}~~$h_{i}=(E_{ii}-{d_{i+1}\over
d_{i}}E_{i+1,i+1}),$\>\>\>\>(2.1)\\[4mm] with
$d_{1}=d_{2}=-d_{3}=-d_{4}=1$,\\[4mm]

{}~~~~b) the Serre-relations:\\[2mm]
\>\>~~$[E_{12},E_{34}]$\>=\>$[E_{21},E_{43}]$ \>
{}~~~~~~~=~
0,\\[1mm]
\>\>~~~~~~$E_{23}^{2}$\>=\>~~~~$E_{32}^{2}$\>
{}~~~~~~~=~
0,\\[1mm]
\>\>~~$[E_{12},E_{13}]_{q^{2}}$\>=\>$[E_{24},
E_{34}]_{q^{2}}$
\>~~~~~~~=~0,\\[1mm]
\>\>~~$[E_{21},E_{31}]_{q^{2}}$\>=\>$[E_{42},
E_{43}]_{q^{2}}$
\>~~~~~~~=~0,
\>\>\>(2.2)\\[2mm]
and\\[2mm]

{}~~~~c) the extra-Serre relations $^{12}$:\\[2mm]
\>\>~~~~$\{E_{13},E_{24}\}$\>=\>0,\\[1mm]
\>\>~~~~$\{E_{31},E_{42}\}$\>=\>0,\>\>\>\> (2.3)\\[2mm]
where the operators
\\[2mm]
\>\>~~~~~~~~~~~~$E_{13}$~:\>=\>$[E_{12},
E_{23}]_{q^{-2}}$, \\[1mm]
\>\>~~~~~~~~~~~~$E_{24}$~:\>=\>$[E_{23},
E_{34}]_{q^{-2}}$, \\[1mm]
\>\>~~~~~~~~~~~~$E_{31}$~:\>=\>$-[E_{21},
E_{32}]_{q^{-2}},$ \\[1mm]
\>\>~~~~~~~~~~~~$E_{42}$~:\>=\>$-[E_{32},
E_{43}]_{q^{-2}}$.
\>\>\>\> (2.4)
\end{tabbing}
and the operators composed in the following way
\begin{tabbing}
\=123456789123456789\=$E_{41}$~\=:=1
\=$[E_{21},[E_{32},E_{43}]_{q^{-2}}]_{q^{-2}}$
\=$\equiv [E_{21},E_{42}]_{q^{-1}}$123456789123 \=\kill
\>\>$E_{14}$~\>:=\>$[E_{12},[E_{23},
E_{34}]_{q^{-2}}]_{q^{-2}}$
\>$\equiv ~[E_{12},E_{24}]_{q^{-2}}$, \\[1mm]
\>\>$E_{41}$\>:=\>$[E_{21},[E_{32},
E_{43}]_{q^{-2}}]_{q^{-2}}$
\>$\equiv ~-[E_{21},E_{42}]_{q^{-2}}$ \>(2.5)
\end{tabbing}
were defined in {\bf I} as new generators. The latter
are odd and have vanishing squares. They, together
with the Cartan-Chevalley generators, form a full
system of q-analogues of the Weyl generators $e_{ij}$,
$1\leq i,j\leq 4$,
$$(e_{ij})_{kl} = \delta_{ik}\delta_{jl},
\eqno(2.6)$$
of the superalgebra $gl(2/2)$ whose universal
enveloping algebra $U[gl(2/2)]$ is a classical
limit of $U_{q}[gl(2/2)]$ when $q\rightarrow 1$.
Other commutation relations between $E_{ij}$,
used in different calculations through out {\bf I}
and the present paper, follow from the relations
(2.1)-(2.3) and the definitions (2.4) and (2.5).\\[2mm]

  {\bf A. The induced modules}\\

  Since
$U_{q}[gl(2/2)_{0}]\equiv
U_{q}[gl(2)_{l}\oplus gl(2)_{r}]$
generated by $E_{ij}$, $1\leq i,j\leq 2$ or
$3\leq i,j\leq 4$, is a stability subalgebra
of $U_{q}[gl(2/2)]$ we can construct
finite--dimensional representations of
$U_{q}[gl(2/2)]$ induced from
finite--dimensional representations of
$U_{q}[gl(2/2)_{0}]$ which, as was shown by
Rosso $^{17}$ and Lusztig $^{18}$, are simply
quantum deformations (q-deformations) of
finite--dimensional representations of the
classical $gl(2)\oplus gl(2)$. In {\bf I},
the $U_{q}[gl(2/2)]$--module $W^{q}$
induced from a finite--dimensional
irreducible $U_{q}[gl(2/2)_{0}]$--module
$V_{0}^{q}$ such that
$$E_{23}V_{0}^{q}=0,\eqno(2.7)$$
by the construction, is  the factor--space

$$W^{q}=[U_{q}\otimes V_{0}^{q}]/I^{q},\eqno(2.8)$$
where
$U_{q}:=U_{q}[gl(2/2)]$, while

$$I^{q}={\normalsize lin.env.}\{ub\otimes v -u\otimes bv\|
{}~ u\in U_{q},~ b\in U_{q}(B)\subset U_{q},~ v\in
V_{0}^{q}\} \eqno(2.9)$$
and
$$U_{q}(B)={\normalsize lin.env.}\{E_{ij},E_{23}\|i,j=1,2~~
{\normalsize and}~~i,j=3,4\}.
\eqno(2.10) $$
According to (2.8) any vector $w$ from the induced
module $W^{q}$ has the form $$w=u\otimes v,~~~
u\in U_{q},~~ v\in
V_{0}^{q}.\eqno(2.11)$$
Then $W^{q}$ is a $U_{q}[gl(2/2)]$-module
in the sense $$gw\equiv g(u\otimes v)=gu\otimes v\in
W^{q}\eqno(2.12)$$ for  $g$, $u\in U_{q}$,
$w\in W^{q}$ and $v\in
V_{0}^{q}$.\\

   As explained in {\bf I}, for a (GZ) basis
of the $U_{q}[gl(2/2)_{0}]$-module $V_{0}^{q}$
we can take the tensor product

$$
(m):=\left[
\begin{array}{lccc}

                  m_{13}& m_{23}&  m_{33}& m_{43} \\

                  m_{13}& m_{23}&  m_{33}&  m_{43}\\
m_{11}&   0   &  m_{31}&    0
\end{array}
\right]:=
\left[
\begin{array}{c}
                        m_{13}~~~m_{23}\\ m_{11}
\end{array}
\right]
\otimes
\left[
\begin{array}{c}
                        m_{33}~~~m_{43}\\ m_{31}
\end{array}
\right]
\eqno(2.13)$$
between the GZ basis $(m)_{l}$ of $U_{q}[gl(2)_{l}]$
$$
(m)_{l}:=
\left[
\begin{array}{c}
                            [m]_{l}\\ m_{11}
\end{array}
\right]:=
\left[
\begin{array}{c}
                        m_{13}~~~m_{23}\\ m_{11}
\end{array}
\right]
\eqno(2.14)$$
and the GZ basis $(m)_{r}$ of $U_{q}[gl(2)_{r}]$
$$(m)_{r}:=
\left[
\begin{array}{c}
                            [m]_{r}\\ m_{31}
\end{array}
\right]:=
\left[
\begin{array}{c}
                        m_{33}~~~m_{43}\\ m_{31}
\end{array}
\right],
\eqno(2.15)$$
where $m_{ij}$ are complex numbers satisfying (1.3).
Then the decomposition (1.6) can be written explicitly
as follows $^{1}$
\begin{eqnarray*}
W^{q}([m])&=&
V_{(00)}^{q}([m_{13}, m_{23}, m_{33}, m_{43}])
\\[2mm]
& &\bigoplus_{i=0}^{min(1,2l)}
{}~\bigoplus_{j=0}^{min(1,2l')}
V_{(10)}^{q}([m_{13}-i, m_{23}+i-1, m_{33}-j+1, m_{43}+j])
\\[2mm]
& &\bigoplus_{i=0}^{min(2,2l)}
V_{(11)}^{q}([m_{13}-i, m_{23}+i-2, m_{33}+1, m_{43}+1])
\\[2mm]
& &\bigoplus_{j=0}^{min(2,2l')}
V_{(20)}^{q}([m_{13}-1, m_{23}-1, m_{33}-j+2, m_{43}+j])
\\[2mm]
& &\bigoplus_{i=0}^{min(1,2l)}~\bigoplus_{j=0}^{min(1,2l')}
V_{(21)}^{q}([m_{13}-i-1, m_{23}+i-2, m_{33}-j+2, m_{43}+j+1])
\\[2mm]
& &~~~ \bigoplus
V_{(22)}^{q}([m_{13}-2, m_{23}-2, m_{33}+2, m_{43}+2]),
{}~~~~~~~~~~~~~~~~~~~~~(2.16)
\end{eqnarray*}
where $V^{q}_{(00)} \equiv
V^{q}_{0}$, ~$l={1\over 2}(m_{13}-m_{23})$,
{}~$l'={1\over 2}(m_{33}-m_{43})$.
The basis within the module
$W^{q}([m])=W^{q}([m_{13},m_{23},m_{33},m_{43}])$ is
spanned on all the possible QGZ tableaux
(see {\bf I} and also Ref. 14 and 15)
$$(m)_{(hk)}\equiv \left[
\begin{array}{lccc}

                  m_{13}& m_{23}&  m_{33}& m_{43} \\

                  m_{12}& m_{22}&  m_{32}&  m_{42}\\
m_{11}&   0   &  m_{31}&    0
\end{array}
\right]_{(hk)},~~~~h,k\in \{0,1,2\} \eqno(2.17)$$
such that
\begin{eqnarray*}
{}~~~~~~~~~~~~~~~~~~~~~~~
m_{12} &=& m_{13}-r-\theta (h-2) -
\theta (k-2) +1,\\[1mm]
         m_{22} &=& m_{23}+r-\theta (h-1) -
\theta (k-1) -1,\\[1mm]
	    m_{32} &=& m_{33}+h-s+1,\\[1mm]
         m_{42} &=& m_{43}+k+s-1,
{}~~~~~~~~~~~~~~~~~~~~~~~~~~~~~~~~~~~~~~~~~~~~~ (2.18)
\end{eqnarray*}
where
\begin{eqnarray*}
{}~~~~~~~~~~~~~~~~~~~~~~~~~~
r &=& 1,...,1 + {\normalsize min}(h-k,2l'), \\[1mm]
s &=& 1,...,1 + {\normalsize min}(\left<h\right> +
\left<k\right>,2l),
{}~~~~~~~~~~~~~~~~~~~~~~~~~~~~~ (2.19)
\end{eqnarray*}

$$\theta (x)=\left\{
\begin{array}{ll}
1 & ~~~~ {\normalsize if} ~~ x\geq 0,\\
0 & ~~~~ {\normalsize if} ~~ x < 0
\end{array}\right.
\eqno(2.20)$$
\vspace*{1mm}
and

$$\left<i\right>=\left\{
\begin{array}{ll}
1 & ~~~~ {\normalsize for~~ odd}~~ i,\\
0 & ~~~~ {\normalsize for~~ even}~~ i~.
\end{array}\right.
\eqno(2.21)$$
Now we are ready to write explicitly down the matrix
elements for all the (sufficiently, Cartan--Chevalley)
generators of $U_{q}[gl(2/2)]$.
\\[2mm]

   {\bf B. Typical representations}\\

     When the condition (1.9), i.e., (I.2.41), holds
the finite--dimensional module $W^{q}$ is irreducible and
called typical. In that case, the transformations of the
basis (2.17) of $W^{q}$ under the actions of
the generators of $U_{q}[gl(2/2)]$ are explicitly given
in {\bf I}. After using the formal rule (I.4.44)
$$\left|{[x_{1}]...[x_{i}]\over [y_{1}]...[y_{j}]}\right|:=
{[|x_{1}|]...[|x_{i}|]\over [|y_{1}|]...[|y_{j}|]}\eqno(2.22)$$
assumed in
{\bf I} and removing the modulus, the expressions of the
matrix elements of the even generators of
$U_{q}[gl(2/2)]$
remain the same as in {\bf I} (see (I.4.43)), while
the matrix elements (I.4.45) and (I.4.46) of the generators
$E_{23}$ and $E_{32}$ are reorganized (with a slight
simplification) respectively as follows $^{1}$:
\begin{eqnarray*}
E_{23}(m)_{(00)} &=& 0,\\[2mm]
E_{23}(m)_{(10)} &=&
-[l_{3-r,3} + l_{s+2,3}+3]
\left({(-1)^{s+r}[l_{3-r,3}-l_{11}][l_{5-s,3}-l_{31}+1]
\over
[l_{13}-l_{23}][l_{33}-l_{43}]}\right)^{1/2}
(m)_{(00)}^{+i2-j2-31},\\[4mm]
E_{23}(m)_{(ab)} &=&
\sum_{i=max(1,b-r+2)}^{min(2,b-r+3)}
{}~~\sum_{j=max(3,b+s+1)}^{min(4,b+s+2)}
(-1)^{(b-1)i+b(j+1)}\\[2mm]
& &\times [l_{i3}+l_{j3}+\left<s\right> -
\left<r\right>+3]\\[2mm]
& & \times \left({(-1)^{i+j+1}[l_{i2}-l_{11}+1]
[l_{7-j,2}-l_{31}+1]\over
[l_{12}-l_{22}][l_{32}-l_{42}]}\right)^{1/2}\\[2mm]
& & \times
\left({[l_{i2}-l_{3-i,2}+\left<r\right>-1]\over
[2-\left<r\right>] [l_{i3}-l_{3-i,3}+(-1)^{r}]}
\right)^{b/2}\\[2mm]
& &\times \left({[l_{j2}-l_{7-j,2} -
\left<s\right>+1]\over
[2-\left<s\right>] [l_{j3}-l_{7-j,3}-(-1)^{s}]}
\right)^{(1-b)/2}
(m)_{(10)}^{+i2-j2-31},~~~a+b=2,\\[4mm]
E_{23}(m)_{(21)} &=&
-\sum_{i=1}^{2}\sum_{j=3}^{4}~~
\sum_{\left<s+j\right> \leq k=0,1\leq \left<r+i\right>}
(-1)^{(1-k)i+kj}\\[2mm]
& &\times
[l_{i3}+l_{j3}-(-1)^{k}\left<i+j+s+r\right>+3]\\[2mm]
& &\times \left({(-1)^{i+j+1}[l_{i2}-l_{11}+1]
[l_{7-j,2}-l_{31}+1]\over
[l_{12}-l_{22}][l_{32}-l_{42}]}\right)^{1/2}\\[2mm]
& & \times
\left({(-1)^{r+1}[l_{r2}-l_{i2}+2k-2]\over
[2-k][l_{13}-l_{23}]}\right)^{\left<i+r\right>/2}\\[2mm]
& &\times \left({(-1)^{s}[l_{5-s,2}-l_{j2}+2k]\over
[1+k][l_{33}-l_{43}]}\right)^{\left<s+j+1\right>/2}
(m)_{(1+k,1-k)}^{+i2-j2-31},\\[4mm]
E_{23}(m)_{(22)} &=&
\sum_{i=1}^{2}\sum_{j=3}^{4}
(-1)^{i+j}
[l_{i3}+l_{j3}+3]\\[2mm]
& &\times \left({(-1)^{i+j+1}[l_{i3}-l_{11}-1]
[l_{7-j,3}-l_{31}+3]\over
[l_{13}-l_{23}][l_{33}-l_{43}]}
\right)^{1/2} (m)^{+i2-j2-31}_{(21)},
{}~~~~(2.23)
\end{eqnarray*}
\\[2mm]
and

\begin{eqnarray*}
E_{32}(m)_{(00)} &=& -
\sum_{i=1}^{2}\sum_{j=3}^{4}
  \left({(-1)^{i+j+1}[l_{i3}-l_{11}]
[l_{7-j,3}-l_{31}]\over
  [l_{13}-l_{23}][l_{33}-l_{43}]}\right)^{1/2}
  (m)_{(10)}^{-i2+j2+31},\\[4mm]
  E_{32}(m)_{(10)} &=&
\sum_{i=1}^{2}\sum_{j=3}^{4}
{}~~\sum_{\left<r+i\right>\leq k=0,1\leq
\left<s+j\right>}(-1)^{(1-k)i+k(j+1)}\\[2mm]
& &\times \left({(-1)^{i+j+1}[l_{i2}-l_{11}]
[l_{7-j,2}-l_{31}]\over
  [l_{12}-l_{22}][l_{32}-l_{42}]}\right)^{1/2}\\[2mm]
& &\times
 \left({(-1)^{r}[l_{3-i,3}-l_{i3}+2k-1]\over
[1+k][l_{13}-l_{23}]} \right)^{\left<i+r+1\right>/2}\\[2mm]
& &\times \left({(-1)^{s+1}[l_{7-j,3}-l_{j3}+2k-1]\over
[2-k][l_{33}-l_{43}]}\right)^{\left<s+j\right>/2}
  (m)_{(2-k,k)}^{-i2+j2+31},\\[4mm]
 E_{32}(m)_{(ab)}
  &=& -\sum_{i=max(1,r-b)}^{min(2,r-b+1)}
 ~~\sum_{j=max(3,5-b-s)}^{min(4,6-b-s)}
(-1)^{bi+(1-b)j}\\[2mm]
& &\times \left({(-1)^{i+j+1}[l_{i2}-l_{11}]
[l_{7-j,2}-l_{31}]\over
  [l_{12}-l_{22}][l_{32}-l_{42}]}\right)^{1/2}\\[2mm]
& &\times
 \left({[l_{i2}-l_{3-i,2}-\left<r\right>+1]\over
  [2-\left<r\right>] [l_{i3}-l_{3-i,3}-(-1)^{r}]}
\right)^{b/2}\\[2mm]
& &\times \left({[l_{j2}-l_{7-j,2} +
\left<s\right>-1]\over
  [2-\left<s\right>] [l_{j3}-l_{7-j,3}+(-1)^{s}]}
\right)^{(1-b)/2}
  (m)_{(21)}^{-i2+j2+31},~~~a+b=2,\\[4mm]
  E_{32}(m)_{(21)} &=&
-(-1)^{r+s} \left({(-1)^{r+s}[l_{r3}-l_{11}-1]
[l_{s+2,3}-l_{31}+2]\over
[l_{13}-l_{23}][l_{33}-l_{43}]}\right)^{1/2}\\[2mm]
& &\times
(m)^{-r2+j2+31}_{(22)},~~~~j=5-s,\\[4mm]
 E_{32}(m)_{(22)} &=&0,
{}~~~~~~~~~~~~~~~~~~~~~~~~~~~~~~~~~~~~~~~~~~~~~~~~~~
{}~~~~~~~~~~~~~~~~~~~~~~~~~~~(2.24))
\end{eqnarray*}
where the rescaling
$$
\begin{array}{ccc}
E_{23} ~& \longrightarrow &~ q^{2}E_{23},\\
E_{32} ~& \longrightarrow &~ q^{-2}E_{32},\\
\end{array}
\eqno(2.25))$$
and the rescaling (I.4.47) in Ref. 1 have already
been taken into account. Unless stated otherwise,
hereafter these rescalings will be kept and understood
through out the paper.\\

   If the condition (1.9) (i.e., (I.4.41)) is not
fulfilled, namely,
if for certain values of $i=1,2$ and $j=3,4$
$$l_{i3}+l_{j3}+3 =0;~~
i=1,2~ {\normalsize and}~ j=3,4\eqno(2.26)$$
the representations of $U_{q}[gl(2/2)]$ in $W^{q}$
are no longer irreducible but indecomposible (cf.
Refs. 14,15). As we can see later, there are five
possibilities of occurring the equalities (2.26)
leading to five classes of nontypical
representations of $U_{q}[gl(2/2)]$ which are
irreducible representations extracted from
indecomposible representations in modules $W^{q}$.
In the next section, where our plan in many points
coincides with the one of {\bf I$^{\ast}$}, we shall
classify and consider indecomposible representations
and all the nontypical representations of
$U_{q}[gl(2/2)]$.\\[5mm]

\begin{flushleft}
{\bf III. NONTYPICAL REPRESENTATIONS}
\end{flushleft}
\vspace*{2mm}

    When a module $W^{q}$ is indecomposible, it
contains a maximal invariant subspace $I^{q}_{k}$
and simultaneously the compliment to $I^{q}_{k}$
subspace in $W^{q}$ is not invariant under the
actions of $U_{q}[gl(2/2)]$. However, as mentioned
in the introduction, the factor module
$W^{q}_{k} =W^{q}/I^{k}_{q}$ carries an irreducible
representation of $U_{q}[gl(2/2)]$ which is called
nontypical. It turns out that all the assertions proved
in {\bf I$^{\ast}$} can be extended to take place in
the present case of
the quantum deformation at generic $q$. Because the proofs,
some of which represent direct computations,
are cumbersome, in this section we shall not prove all
these assertions but only some of them.\\

   Since
$m_{13}-m_{23}\in {\bf Z_{+}}$ and
$m_{33}-m_{43}\in {\bf Z_{+}}$
or, equivalently,
$l_{13}-l_{23}\in {\bf N}$ and
$l_{33}-l_{43}\in {\bf N}$
the indecomposible representations and, therefore,
the nontypical representations  of $U_{q}[gl(2/2)]$
are classified, as in the classical case $^{15}$, in
five following classes (see (2.26))

\begin{tabbing}
\=
{}~~~~~~~~~~~~~~~~~~ \= class 1~~~
\= $l_{13} + l_{43} + 3$ \= = 0,
{}~~ \= $\Leftrightarrow$~~~ \= $m_{13} + m_{43} - 1$ = 0
{}~~~~~~~~~~~~~~\=\kill
\>   \> class 1   \>$l_{13} + l_{43} + 3$ \> = 0
\> $\Leftrightarrow$ \> $m_{13} + m_{43}$  = 0, \\[2mm]
\>   \>           \>$l_{23} + l_{33} + 3$ \> $\neq 0$
\> $\Leftrightarrow$ \> $m_{23} + m_{33}$  $\neq 0$;
\> (3.1)\\[4mm]
\>   \> class 2   \>$l_{13} + l_{43} + 3$ \> $\neq 0$
\> $\Leftrightarrow$ \> $m_{13} + m_{43}$  $\neq 0$,\\[2mm]
\>   \>           \>$l_{23} + l_{33} + 3$ \> = 0
\> $\Leftrightarrow$ \> $m_{23} + m_{33}$  = 0;
\> (3.2)\\[4mm]
\>   \> class 3   \>$l_{23} + l_{43} + 3$ \> = 0
\> $\Leftrightarrow$ \> $m_{23} + m_{43}  - 1$  = 0;
\> (3.3)\\[4mm]
\>   \> class 4   \>$l_{13} + l_{33} + 3$ \> = 0
\> $\Leftrightarrow$ \> $m_{13} + m_{33} + 1$  = 0;
\> (3.4)\\[4mm]
\>   \> class 5   \>$l_{13} + l_{43} + 3$ \> = 0
\> $\Leftrightarrow$ \> $m_{13} + m_{43}$  = 0,\\[2mm]
\>   \>           \>$l_{23} + l_{33} + 3$ \> = 0
\> $\Leftrightarrow$ \> $m_{23} + m_{33}$  = 0. \> (3.5)
\end{tabbing}

     Let us consider an induced module
$W^{q}([m])$,~$[m]=[m_{13},m_{23},m_{33},m_{43}]$,
of $U_{q}[gl(2/2)]$.
According to (2.16) the module $W^{q}([m])$ is decomposed
into a sum of all possible 16 or less
$U_{q}[gl(2)\oplus gl(2)]$ - fidimods
$V_{(hk)}([m]_{(hk)})$, where $h,k\in \{0,1,2\}$, while
the signature
$[m]_{(hk)}\equiv [m_{12},m_{22}, m_{32},m_{42}]_{(hk)}$
determined by the same formula (and also by (2.18)-(2.21))
has the form
$$[m_{12},m_{22},m_{32},m_{42}]_{(hk)} =
[m_{13}-a ,m_{23}-b ,
m_{33}+c ,m_{43}+d]_{(hk)},
{}~~ a,b^,c,d \in {\bf Z_{+}}.
\eqno(3.6)$$
The subscript $(hk)$ can be omitted when there is no a
confusion caused by some degenerations. As in {\bf I$^{\ast}$},
we assume the notation (with a slight modification)
$$V^{q}_{(s)}([m_{13}-a ,m_{23}-b, m_{33}+c ,m_{43}+d]):=
{\frame{\shortstack{{\small ~-a,-b,c,d}}~}}_{~(s)}
\eqno(3.7)$$
for an arbitrary
$V^{q}_{(s)}([m_{13}-a ,m_{23}-b, m_{33}+c ,m_{43}+d])$
which is denoted by
$$V^{q}_{(s)}([m_{13}-a ,m_{23}-b, m_{33}+c ,m_{43}+d]):=
{}~~\stackrel{{\small inv}}
{\frame{\shortstack{
{\small ~-a,-b,c,d}}~}}_{~(s)}
\eqno(3.8)$$
if
$$V^{q}_{(s)}([m_{13}-a ,m_{23}-b, m_{33}+c ,m_{43}+d])
\subseteq I_{k}^{q}$$
or by
$$V^{q}_{(s)}([m_{13}-a ,m_{23}-b, m_{33}+c ,m_{43}+d]):=
{}~~\stackrel{{\small noninv}}
{\frame{\shortstack{
{\small ~-a,-b,c,d}}~}}_{~(s)}
\eqno(3.9)$$
if otherwise,
$$V^{q}_{(s)}([m_{13}-a ,m_{23}-b, m_{33}+c ,m_{43}+d])
\cap I_{k}^{q} = \O,$$
where by $(s)$ we denote $(hk)$, upper, lower or any other
indices specifying the subspaces $V^{q}_{(s)}$. As in the
classical case $^{15}$, for some $V^{q}_{(s)}$ neither (3.8)
nor (3.9) holds.\\

   As mentioned above, although the matrix elements and
other expressions are deformed all the propositions proved
in {\bf I$^{\ast}$} have in the present case quantum analogues.
Bellow, we shall prove some of the most important
Propositions.\\

   {\it Proposition} 1: For any $k=1,2,3,4,5$ the submodule
$V^{q}_{(22)}([m_{13}-2 ,m_{23}-2, m_{33}+2 ,m_{43}+2])$
always belongs to the maximal
$U_{q}[gl(2/2)]$-invariant subspace $I^{q}_{k}$ of
the class $k$ indecomposible
induced module $W^{q}$
$$V^{q}_{(22)}([m_{13}-2 ,m_{23}-2, m_{33}+2 ,m_{43}+2])
=~~\stackrel{{\small inv}}
{\frame{\shortstack{
{\small ~-2,-2,2,2}}~}}_{~(22)}~ \subset I^{q}_{k}.
\eqno(3.10)$$

   {\it Proof}: This proposition is a quantum analogue
of Proposition 4 in {\bf I$^{\ast}$}. Following the latter we
suppose
$$0\neq x\in I_{k}^{q}.\eqno(3.11)$$
Then from {\bf I} we have
$$x=\sum_{\theta_{i}=0,1}~~ \sum_{(m)\in V_{(0)}^{q}}\alpha
\left(\theta_{1}, \theta_{2},
\theta_{3}, \theta_{4};(m)\right)
(E_{31})^{\theta_{1}} (E_{32})^{\theta_{2}}
(E_{41})^{\theta_{3}} (E_{42})^{\theta_{4}}\otimes (m).
\eqno(3.12)$$
Let $k$ be a number (which is never bigger than 4) such
that if
{}~$\sum_{i=1}^{4}\theta_{i} < k$
all the coefficients
{}~$\alpha \left(\theta_{1}, \theta_{2},
\theta_{3}, \theta_{4};(m)\right)=0$  and
for certain $(m^{0})$ and $\theta_{i}^{0}$,
{}~ $\sum_{i=1}^{4}\theta_{i}^{0} = k$,
{}~$\alpha \left(\theta_{1}^{0}, \theta_{2}^{0},
\theta_{3}^{0}, \theta_{4}^{0};(m^{0})\right) \neq 0$.
Thus the first sum in (3.12) is truncated and spreads
over only all $\theta_{i}$, $i=1,2,3,4$, for which
$\sum_{i=1}^{4}\theta_{i}\geq k$. From (3.11)
we have also
$$0\neq
(E_{31})^{1-\theta^{0}_{1}} (E_{32})^{1-\theta^{0}_{2}}
(E_{41})^{1-\theta^{0}_{3}} (E_{42})^{1-\theta^{0}_{4}}x
:= y
\in I_{k}^{q}.\eqno(3.13)$$
 Using (2-1)--(2.5) (or (I.3.1)-(I.3.5)), (I.4.33) and (I.4.34) (i.e.,
(2.16)) we derive
$$(E_{31})^{1-\theta^{0}_{1}} (E_{32})^{1-\theta^{0}_{2}}
(E_{41})^{1-\theta^{0}_{3}} (E_{42})^{1-\theta^{0}_{4}}
(E_{31})^{\theta^{0}_{1}} (E_{32})^{\theta^{0}_{2}}
(E_{41})^{\theta^{0}_{3}} (E_{42})^{\theta^{0}_{4}}$$
$$=(-1)^{(1 - \theta^{0}_{4})(\theta^{0}_{1} +
\theta^{0}_{2} + \theta^{0}_{3})
+ (1 - \theta^{0}_{3})(\theta^{0}_{1} + \theta^{0}_{2}) +
(1 - \theta^{0}_{2})\theta^{0}_{1}}
E_{31} E_{32} E_{41} E_{42}\eqno(3.14)$$
and according to (I.4.33) and (2.16) (or (I.4.34))

$$y\in E_{31} E_{32} E_{41} E_{42}\otimes V_{0}^{q}([m])
\equiv
V_{22}^{q}([m_{13}-2, m_{23}-2, m_{33}-2, m_{43}-2]).
\eqno(3.14)$$
In other words,
$$V^{q}_{(22)}([m_{13}-2 ,m_{23}-2, m_{33}+2 ,m_{43}+2])
\subset I_{k}^{q},$$
since $V^{q}_{(22)}$ is a
$U_{q}[gl(2) \oplus gl(2)]$-fidirmod
and $I_{k}^{q}$ is the maximal $U_{q}[gl(2/2)]$ invariant
subspace of the indecomposible module $W^{q}$. Therefore
(3.10) holds.\\

  Note that for a proof of the latter proposition
 we can choose, for example,
$$x=\sum_{\eta_{i}=0,1}~~ \sum_{(m)\in V_{(0)}^{q}}\alpha
\left(\eta_{1}, \eta_{2},
\eta_{3}, \eta_{4};(m)\right)
(E_{41})^{\eta_{1}} (E_{31})^{\eta_{2}}
(E_{42})^{\eta_{3}} (E_{32})^{\eta_{4}}\otimes (m)$$
\vspace*{2mm}
instead (3.12).\\

    The following proposition is a quantum
analogue of Proposition 5 in {\bf I$^{\ast}$}: \\

  {\it Proposition} 2: For any $k=1,2,3,4,5$
$$V^{q}_{(00)}([m_{13}, m_{23}, m_{33}, m_{43}])
= ~~\stackrel{{\small noninv}}
{\frame{
{\small ~0,0,0,0}~}}_{~(00)}\cap I_{k}^{q} = \O.
%%{\frame{\shortstack{
%%{\small ~0,0,0,0}}~}}_{~(00)}\cap ~I_{k}^{q} = \O.
\eqno(3.15)$$

   {\it Proof}: If
$$V^{q}_{(00)}([m_{13}, m_{23}, m_{33}, m_{43}])
\cap I_{k}^{q} \neq \O$$
then
$$V^{q}_{(00)}([m_{13}, m_{23}, m_{33}, m_{43}])
\subset I_{k}^{q}.$$
Therefore, according to (I.4.19$^{'}$) and the fact that
the proper subspace $I_{k}^{q}$ of $W^{q}$ is
invariant under the actions of $U_{q}[gl(2/2)]$ we
would have  the contradiction
$$W^{q}=T^{q} \odot V_{(00)}^{q} \subset I_{k}^{q}$$
which means  $I_{k}^{q}$ ($=W^{q}$) is not a proper
subspace of $W^{q}$ as it has to be.\\

   We are also in a position to prove the following
assertion:\\

   {\it Proposition} 3: The transformation of the factor
space $W^{q}_{k}([m])\equiv W^{q}([m])/I^{q}_{k}$ under
the action of $U_{q}[gl(2/2)]$ can be obtained by replacing
in the corresponding transformation of $W^{q}([m])$
all the basis vectors belonging to the maximal invariant
subspace $I^{q}_{k}$ by zero.\\

      The proof of this proposition is analogous with
the one of Proposition 6 in {\bf I$^{\ast}$}.\\

   Let us now consider the nontypical representations of
$U_{q}[gl(2/2)]$ of all the classes corresponding to the
conditions (3.1)-(3.5). Since the decomposition structures
of the (quantum) indecomposible and nontypical modules are
the same as those of the classical ones, bellow, in order
to make the present paper less cumbersome we shall expose
as examples only some of their decomposition schemes but
not all which can be found in {\bf I$^{\ast}$}. At the meantime
the matrix elements for the nontypical representations of
all the classes are given explicitly in the next subsections
or in the Appendix. A more detailed description will be
made only for the class 1 nontypical representations.\\[2mm]

{\bf A. The class 1 nontypical representations}\\

   This class corresponds to the case (3.1)
\begin{tabbing}
\=
{}~~~~~~~~~~~~~~~~~~ \= class 1~~~ \= $l_{13} + l_{43} + 3$ \= = 0,
{}~~ \= $\Leftrightarrow$~~~ \= $m_{13} + m_{43} - 1$ = 0
{}~~~~~~~~~~~~\=\kill
\>   \> class 1   \>$l_{13} + l_{43} + 3$ \> = 0
\> $\Leftrightarrow$ \> $m_{13} + m_{43}$  = 0, \\[2mm]
\>   \>           \>$l_{23} + l_{33} + 3$ \> $\neq 0$
\> $\Leftrightarrow$ \> $m_{23} + m_{33}$  $\neq 0$.
\> (3.1)\\[4mm]
\end{tabbing}
Then we have to replace everywhere $m_{43}$ with $-m_{13}$
$$W^{q}([m;~ m_{43}=-m_{13}])
:=W^{q}([m_{13}, m_{23}, m_{33}, -m_{13}])
\eqno(3.16)$$
and keep valid the conditions
\begin{tabbing}
\=
{}~~~~~~~~~~~~~~~~~~~~~~~~~~~~ \= $m_{13} > +m_{23}$, \=
{}~~ \= $\Leftrightarrow$~~~~ \= $m_{13} + m_{43} - 1$ = 0
{}~~~~~~~~~~~~~~~~~~\=\kill
\>      \>$m_{13} > m_{23}$ \>
\> $\Leftrightarrow$ \> $l_{13} - l_{23} -1 > 0$, \\[2mm]
\>             \>$m_{33} > -m_{13}$ \>
\> $\Leftrightarrow$ \> $l_{13} + l_{33} + 2 >0$
\> (3.17)\\
\end{tabbing}
through out this subsection. Once these conditions are
violated for some vectors we have to put the latter equal
zero.\\

   As mentioned above, there are some subspaces
$V^{q}_{(s)}$, namely, the submodules
$V^{q}_{(11)}([m_{13}-1,m_{22}-1,m_{32}+1,m_{42}+1])$
{}~and~
$V^{q}_{(20)}([m_{13}-1,m_{22}-1,m_{32}+1,m_{42}+1])$,
for which
neither (3.8) nor (3.9) holds. However, we can find
their linear combinations satisfying either (3.8) or
(3.9). Indeed, being $U_{q}[gl(2)\oplus gl(2)]$ -
fidirmods with a signature
$$[m\pm 1;~ m_{43}=-m_{13}]
:= [m_{13}-1, m_{23}-1, m_{33}+1, -m_{13}+1]
\eqno(3.18)$$
the following linear spaces
\begin{eqnarray*}
{}~~~~~~~~\stackrel{{\small inv}}{V}_{(1)}^{~q} & := &
\stackrel{{\small inv}}{V}^{~q}_{(1)}
([m\pm 1;~ m_{43}=-m_{13}])\\
& =  & {\normalsize lin. env.}\{
\stackrel{{\small inv}}{(m_{11}, m_{31})}_{(1)} \|
{}~ m_{13} - m_{11} - 1, m_{11} - m_{23} + 1,\\
& &
m_{33} - m_{31} + 1, m_{31} + m_{13} - 1 \in
{\bf Z}_{+}\}
{}~~~~~~~~~~~~~~~~~~~~~~~~~~~~~~~~~(3.19)
\end{eqnarray*}
and
\begin{eqnarray*}
{}~~~~~~~~V^{q}_{(1)} & := &
V^{q}_{(1)}([m\pm 1;~ m_{43}=-m_{13}])\\
& =  &{\normalsize lin. env.}\{(m_{11}, m_{31})_{(1)} \|
{}~m_{13} - m_{11} - 1, m_{11} - m_{23} + 1,\\
& &
m_{33} - m_{31} + 1, m_{31} + m_{13} - 1 \in
{\bf Z}_{+}\},
{}~~~~~~~~~~~~~~~~~~~~~~~~~~~~~~~~(3.20)
\end{eqnarray*}
belong always to the invariant subspace
$I^{q}_{k}$ and the compliment to it subspace
of $W^{q}$, respectively, where
\begin{eqnarray*}
{}~~~~~~~~~~~~\stackrel{{\small inv}}
{(m_{11}, m_{31})}_{(1)}
& = &
{[2]\over 2}\left\{
\left({[l_{13} - l_{23} + 1]}\over
{[l_{13} - l_{23} - 1]} \right)^{1/2}
(m;~m_{43}=-m_{13})_{(11)}^{-13-23+33+43}
\right. \\[4mm]
& & \left.
+\left({[l_{13} + l_{33} + 4]}\over
{[l_{13} + l_{33} + 2]} \right)^{1/2}
(m;~m_{43}=-m_{13})_{(20)}^{-13-23+33+43} \right\}.
\\& &~~~~~~~~~~~~~~~~~~~~~~~~~~~~~~
{}~~ ~ ~ ~~  ~ ~ ~~ ~ ~ ~ ~ ~ ~ ~ ~ ~ ~ ~ ~ ~
{}~~ ~ ~ ~~ ~~~ ~~ ~
(3.21)
\end{eqnarray*}
and\\[2mm]
\begin{eqnarray*}
{}~~~~~~~~~~~~(m_{11}, m_{31})_{(1)} & := &
(m;~m_{43}=-m_{13})_{(1)}^{-13-23+33+43}
\\[4mm]
& = & {[2]\over 2}\left\{
\left({[l_{13} - l_{23} + 1]}\over
{[l_{13} - l_{23} - 1]} \right)^{1/2}
(m;~m_{43}=-m_{13})_{(11)}^{-13-23+33+43}
\right. \\[4mm]
& & \left.
 -\left({[l_{13} + l_{33} + 4]}\over
{[l_{13} + l_{33} + 2]} \right)^{1/2}
(m;~m_{43}=-m_{13})_{(20)}^{-13-23+33+43} \right\}.
\\& &~~~~~~~~~~~~~~~~~~~~~~~~~~~~~~
{}~~ ~ ~ ~~  ~ ~ ~~ ~ ~ ~ ~ ~ ~ ~ ~ ~ ~ ~ ~ ~
{}~~ ~ ~ ~~ ~~~ ~~ ~
(3.22)
\end{eqnarray*}
In (3.21) and (3.22) and hereafter by
{}~$(m;~m_{kl}=\alpha )_{(s)}^{\pm ij}$~
we denote a QGZ basis vector obtained from ~$(m)$~
by replacing the elements ~$m_{kl}$~ and ~$m_{ij}$~ of
the latter by ~$\alpha$~ and ~$m_{ij}\pm 1$~, respectively,
while ~$[m;~m_{kl}=\alpha]^{\pm ij}$~ is obtained from the
signature ~$[m]$~ by the same way. The index $(s)$ indicates
the subspace to which the considered vector belongs.\\

   {\it Proposition} 4:  Let $I^{q}_{1}$ be a
maximal invariant subspace in a class 1
indecomposible induced
$U_{q}[gl(2/2)]$-module $W^{q}([m;~m_{43}=-m_{13}])$,
then
$$\stackrel{{\small inv}}
{V}^{~q}_{(1)}([m\pm 1;~m_{43}=-m_{13}])
=~~\stackrel{{\small inv}}
{\frame{\shortstack{
{\small ~-1,-1,1,1}}~}}_{~(1)},
\eqno(3.23)$$
i.e.,
$$\stackrel{{\small inv}}
{V}^{~q}_{(1)}([m\pm 1;~m_{43}=-m_{13}])
{}~\subset I^{q}_{1}$$
and
$$V^{q}_{(1)}([m\pm 1;~m_{43}=-m_{13}])
=~~\stackrel{{\small noninv}}
{\frame{\shortstack{
{\small ~-1,-1,1,1}}~}}_{~(1)},
\eqno(3.24)$$
i.e.,
$$V^{q}_{(1)}([m\pm 1;~m_{43}=-m_{13}])
{}~\cap I^{q}_{1} =\O.$$
The module $W^{q}([m;~m_{43}=-m_{13}])$ is
decomposed exactly
in the same way as the classical module
$W([m;~m_{43}=-m_{13}])$
(see {\bf I$^{\ast}$})
$$
\begin{array}{cclclcl}
W^{q}([m;~m_{43}=-m_{13}])~~
=& &\stackrel{{\small noninv}}
{\frame{\shortstack{
{\small ~0,0,0,0}}~}}_{~(00)}&\oplus &
\stackrel{{\small noninv}}
{\frame{\shortstack{
{\small ~-1,0,1,0}}~}}_{~(10)}&\oplus &
\stackrel{{\small noninv}}
{\frame{\shortstack{
{\small ~0,-1,1,0}}~}}_{~(10)}\\
&\oplus &
\stackrel{{\small inv}}
{\frame{\shortstack{
{\small ~-1,0,0,1}}~}}_{~(10)}
&\oplus &
\stackrel{{\small noninv}}
{\frame{\shortstack{
{\small ~0,-1,0,1}}~}}_{~(10)}&\oplus &
\stackrel{{\small noninv}}
{\frame{\shortstack{
{\small ~-1,-1,2,0}}~}}_{~(20)}\\
&\oplus&
\stackrel{{\small inv}}
{\frame{\shortstack{
{\small ~-1,-1,0,2}}~}}_{~(20)}&\oplus &
\stackrel{{\small inv}}
{\frame{\shortstack{
{\small ~-2,0,1,1}}~}}_{~(11)}
&\oplus &
\stackrel{{\small noninv}}
{\frame{\shortstack{
{\small ~0,-2,1,1}}~}}_{~(11)}\\
&\oplus &
\stackrel{{\small inv}}
{\frame{\shortstack{
{\small ~-1,-1,1,1}}~}}_{~(1)}&\oplus &
\stackrel{{\small noninv}}
{\frame{\shortstack{
{\small ~-1,-1,1,1}}~}}_{~(1)}
&\oplus &
\stackrel{{\small inv}}
{\frame{\shortstack{
{\small ~-2,-1,2,1}}~}}_{~(21)}\\
&\oplus &
\stackrel{{\small noninv}}
{\frame{\shortstack{
{\small ~-1,-2,2,1}}~}}_{~(21)}
&\oplus &
\stackrel{{\small inv}}
{\frame{\shortstack{
{\small ~-2,-1,1,2}}~}}_{~(21)}&\oplus &
\stackrel{{\small inv}}
{\frame{\shortstack{
{\small ~-1,-2,1,2}}~}}_{~(21)}\\
&\oplus &
\stackrel{{\small inv}}
{\frame{\shortstack{
{\small ~-2,-2,2,2}}~}}_{~(22)}.
\end{array}
\eqno(3.25)$$
The maximal invariant  subspace ~$I^{q}_{~k}$~ is  a sum
of all the terms
{}~$\stackrel{{\small inv}}
{\frame{\shortstack{{\small ~-a,-b,c,d}}~}}$~ in (3.25)
and represents an irreducible
{}~$U_{q}[gl(2) \oplus gl(2)]$ module with a signature
{}~$[m_{13} - 1, m_{22}, m_{32}, -m_{13}+1]$, while
the compliment to $I^{q}_{~k}$ subspace $W^{q}_{1}$
is a sum of all the terms remaining there, i.e., of all
the terms ~$\stackrel{{\small noninv}}
{\frame{\shortstack{{\small ~-a,-b,c,d}}~}}$ .\\

    Following, in the general line, {\bf I$^{\ast}$},
the proof of the latest proposition (and the proofs of all
the similar propositions for the next classes which often
represent direct computations) is of a rather technical
nature and, therefore, can be omitted.\\

    As in the classical case $^{15}$, the transformations of
the nontypical modules $W^{q}_{1}$ under the actions
of ~$U_{q}[gl(2/2)]$ (more precisely, of its
Cartan-Chevalley generators) are obtained from (2.23) and
(2.24) (or (I.4.45) and (I.4.46)) by\\

    (3.A.1) inserting everywhere there ~$m_{43}=-m_{13}$~
(see (3.1)),\\

    (3.A.2) expressing the vectors
{}~$(m)_{(11)}^{-13-23+33+43}$~ and
{}~$(m)_{(20)}^{-13-23+33+43}$~
in terms of the vectors
{}~$(m_{11}, m_{31})_{(1)}$~ and
{}~$\stackrel{{\small inv}}{(m_{11}, m_{31})}_{(1)}$~
(see (3.21) and (3.22)),\\

    (3.A.3) taking into account Proposition 3, i.e.
replacing all the basis vectors from the maximal invariant
subspace by zero.\\[2mm]

    The actions of the even generators in the nontypical
module  $W^{q}_{1}([m\pm 1;~m_{43}=-m_{13}])$ which is a
direct sum of
irreducible ~$U_{q}[gl(2) \oplus gl(2)]$~ submodules are
standard and given in (I.4.43). As far as the matrix
elements of the odd generators $E_{32}$ and $E_{23}$ are
concerned, in spite of the deformations they have the same
structure with the classical ones and are the following:\\

   Transformations under the action of $E_{32}$:

\begin{eqnarray*}
& &E_{32}(m;~m_{43}=-m_{13})_{(00)}=\\[4mm] & &~~~~
-\left({[l_{13} - l_{11}] [l_{13} + l_{31} + 3] \over
[l_{13} - l_{23}] [l_{13} + l_{33} + 3]} \right)^{1/2}
(m;~m_{43}=-m_{13})_{(10)}^{-13+33+31}\\[4mm]
& &~~~~-\left({[l_{11} - l_{23}] [l_{13} + l_{31} + 3] \over
[l_{13} - l_{23}] [l_{13} + l_{33} + 3]} \right)^{1/2}
(m;~m_{43}=-m_{13})_{(10)}^{-23+33+31}\\[4mm]
& &~~~~-\left({[l_{11} - l_{23}] [l_{33} - l_{31}] \over
[l_{13} - l_{23}] [l_{13} + l_{33} + 3]} \right)^{1/2}
(m;~m_{43}=-m_{13})_{(10)}^{-23+43+31},\\[6mm]
& &E_{32}(m;~m_{43}=-m_{13})_{(10)}^{-13+33}=
\\[4mm] & &~~~~
\left({[l_{11} - l_{23}] [l_{13} + l_{31} + 3] \over
[l_{13} - l_{23}] [l_{13} + l_{33} + 4]}
\right)^{1/2}
(m;~m_{43}=-m_{13})_{(20)}^{-13-23+33+33+31}\\[4mm]
& &~~~~-{1 \over 2}\left({[2][l_{11} - l_{23}]
[l_{33} - l_{31} + 1]
[l_{13} + l_{33} + 3] \over
[l_{13} - l_{23}] [l_{13} + l_{33} + 4]
[l_{13} + l_{33} + 4]}
\right)^{1/2}\\[4mm]
& &~~~~\times
(m;~m_{43}=-m_{13})_{(1)}^{-13-23+33+43+31},\\[6mm]
& &E_{32}(m;~m_{43}=-m_{13})_{(10)}^{-23+33}=
\\[4mm] & &~~~~
-\left({[l_{13} - l_{11}] [l_{13} + l_{31} + 3] \over
[l_{13} - l_{23}] [l_{13} + l_{33} + 4]}
\right)^{1/2}
(m;~m_{43}=-m_{13})_{(20)}^{-13-23+33+33+31}\\[4mm]
& &~~~~-\left({[l_{11} - l_{23} + 1] [l_{33} - l_{31} + 1] \over
[l_{13} - l_{23}+1] [l_{13} + l_{33} + 3]}
\right)^{1/2}
(m;~m_{43}=-m_{13})_{(11)}^{-23-23+33+43+31}\\[4mm]
& &~~~~+{[l_{23} + l_{33} + 3] \over
[l_{13} + l_{33} + 4][l_{13} - l_{23} + 1]}
\left({[l_{13} - l_{11}] [l_{33} - l_{31}+1] \over
[2][l_{13} - l_{23}] [l_{13} + l_{33} + 3]}
\right)^{1/2}\\[4mm]
& &~~~~\times
(m;~m_{43}=-m_{13})_{(1)}^{-13-23+33+43+31},\\[6mm]
& &E_{32}(m;~m_{43}=-m_{13})_{(10)}^{-23+43}=
\\[4mm] & &~~~~
\left({[l_{11} - l_{23} + 1] [l_{13} + l_{31} + 2] \over
[l_{13} - l_{23}+1] [l_{13} + l_{33} + 3]}
\right)^{1/2}
(m;~m_{43}=-m_{13})_{(11)}^{-23-23+33+43+31}\\[4mm]
& &~~~~+ {1 \over [l_{13} - l_{23}+1]}
\left({[l_{13} - l_{11}] [l_{13} - l_{23}]
[l_{13} + l_{31} + 2] \over
[2][l_{13} + l_{33} + 3]}
\right)^{1/2}\\[4mm]
& &~~~~\times
(m;~m_{43}=-m_{13})_{(1)}^{-13-23+33+43+31},\\[6mm]
& &E_{32}(m;~m_{43}=-m_{13})_{(20)}^{-13-23+33+33}=
\\[4mm] & &~~~~
-\left({[l_{11} - l_{23} + 1] [l_{33} - l_{31} + 2]
\over
[l_{13} - l_{23}] [l_{13} + l_{33} + 4]}
\right)^{1/2}
(m;~m_{43}=-m_{13})_{(21)}^{-13-23-23+33+33+43+31},\\[6mm]
& &E_{32}(m;~m_{43}=-m_{13})_{(1)}^{-13-23+33+43}=
\\[4mm] & &~~~~
-\left({[2][l_{11} - l_{23}+1] [l_{13} + l_{31} + 2]
\over
[l_{13} - l_{23}] [l_{13} + l_{33} + 3]}
\right)^{1/2}
(m;~m_{43}=-m_{13})_{(21)}^{-13-23-23+33+33+43+31},\\[6mm]
& &E_{32}(m;~m_{43}=-m_{13})_{(11)}^{-23-23+33+43}=
\\[4mm] & &~~~~
\left({[l_{13} - l_{11}] [l_{13} + l_{31} + 2] \over
[l_{13} - l_{23}+1] [l_{13} + l_{33} + 3]}
\right)^{1/2}
(m;~m_{43}=-m_{13})_{(21)}^{-13-23-23+33+33+43+31},\\[6mm]
& &E_{32}(m;~m_{43}
=-m_{13})_{(21)}^{-13-23-23+33+33+43} = 0.\\
& &~~~~~~~~ ~~~ ~~~ ~~~ ~~ ~~~ ~~~ ~~ ~
{}~~~ ~~ ~~~ ~~~ ~~~ ~~~~ ~~~ ~~ ~~ ~~ ~~ ~~ ~~ ~~ ~~
{}~~ ~~~~ ~~ ~~ ~~ ~~ ~ ~~ ~~ ~~ ~~ ~~ ~~ ~~ ~~ ~~ (3.26)
\end{eqnarray*}

Transformations under the action of $E_{23}$:

\begin{eqnarray*}
& &E_{23}(m;~m_{43}=-m_{13})_{(21)}^{-13-23-23+33+33+43}
=\\[4mm] & &~~~~
\left({[l_{13} - l_{11}] [l_{13} + l_{31} + 1]
[l_{13} + l_{33} + 3] \over
[l_{13} - l_{23}+1]}
\right)^{1/2}
(m;~m_{43}=-m_{13})_{(11)}^{-23-23++33+43-31}\\[4mm]
& &~~~~
+\left({[l_{11} - l_{23} + 1] [l_{33} - l_{31} + 3]
[l_{13} - l_{23}] \over
[l_{13} + l_{33} + 4]}
\right)^{1/2}
(m;~m_{43}=-m_{13})_{(20)}^{-13-23+33+33-31}\\[4mm]
& &~~~~
-[l_{23} + l_{33}+3]
{\left([2][l_{11} - l_{23} + 1] [l_{13} + l_{31} + 1]
[l_{13} - l_{23}] [l_{13} + l_{33} + 3]
\right)^{1/2} \over
[2][l_{13} - l_{23} + 1][l_{13} + l_{33} + 4]}
\\[4mm]
& &~~~~\times
(m;~m_{43}=-m_{13})_{(1)}^{-13-23+33+43-31}
,\\[6mm]
& &E_{23}(m;~m_{43}=-m_{13})_{(11)}^{-23-23+33+43}
=\\[4mm] & &~~~~
[l_{13} - l_{23}]
\left({[l_{11} - l_{23} + 1] [l_{33} - l_{31} + 2]
\over
[l_{13} - l_{23} + 1] [l_{13} + l_{33} + 3]}
\right)^{1/2}
(m;~m_{43}=-m_{13})_{(10)}^{-23++33-31}\\[4mm]
& &~~~~
+[l_{23} + l_{33}+3]
\left({[l_{11} - l_{23} + 1] [l_{13} + l_{31} + 1]
\over
[l_{13} - l_{23} + 1] [l_{13} + l_{23} + 3]}
\right)^{1/2}
(m;~m_{43}=-m_{13})_{(10)}^{-23+43-31}
,\\[6mm]
& &E_{23}(m;~m_{43}=-m_{13})_{(1)}^{-13-23+33+43}
=\\[4mm] & &~~~~
-\left({[2][l_{13} - l_{11}] [l_{33} - l_{31} + 2]
\over [l_{13} - l_{23}] [l_{13} + l_{33} + 3]}
\right)^{1/2}
(m;~m_{43}=-m_{13})_{(10)}^{-23+33-31}\\[4mm]
& &~~~~
+2[l_{13} - l_{23} + 1]
\left({[l_{11} - l_{23}] [l_{33} - l_{31} + 2]
 \over [2][l_{13} + l_{33}+3][l_{13} - l_{23}]}
\right)^{1/2}
(m;~m_{43}=-m_{13})_{(10)}^{-13+33-31}\\[4mm]
& &~~~~
+2[l_{13} + l_{33}+4]
\left({[l_{13} - l_{11}] [l_{13} + l_{31} + 1]
\over
[2][l_{13} - l_{23}][l_{13} + l_{33}+3]}
\right)^{1/2}
(m;~m_{43}=-m_{13})_{(10)}^{-23+43-31}
,\\[6mm]
& &E_{23}(m;~m_{43}=-m_{13})_{(20)}^{-13-23+33+33}
=\\[4mm] & &~~~~
-[l_{13} + l_{33}+3]
\left({[l_{13} - l_{11}] [l_{13} + l_{31} + 2]
\over [l_{13} - l_{23}] [l_{13} + l_{33} + 4]}
\right)^{1/2}
(m;~m_{43}=-m_{13})_{(10)}^{-23+33-31}\\[4mm]
& &~~~~
+[l_{23} + l_{33} + 3]
\left({[l_{11} - l_{23}] [l_{13} + l_{31} + 2]
 \over [l_{13} - l_{23}] [l_{13} + l_{33}+4]}
\right)^{1/2}
(m;~m_{43}=-m_{13})_{(10)}^{-13+33-31}
,\\[6mm]
& &E_{23}(m;~m_{43}=-m_{13})_{(10)}^{-23+43}
=\\[4mm] & &~~~~
\left({[l_{13} - l_{23}] [l_{11} - l_{23}]
[l_{33} - l_{31} + 1]
\over [l_{13} + l_{33} + 3]}
\right)^{1/2}
(m;~m_{43}=-m_{13})_{(00)}^{-31}
,\\[6mm]
& &E_{23}(m;~m_{43}=-m_{13})_{(10)}^{-23+33}
=\\[4mm] & &~~~~
-[l_{23} + l_{33}+3]
\left({[l_{11} - l_{23}] [l_{13} + l_{31} + 2]
\over [l_{13} - l_{23}] [l_{13} + l_{33} + 3]}
\right)^{1/2}
(m;~m_{43}=-m_{13})_{(00)}^{-31}
,\\[6mm]
& &E_{23}(m;~m_{43}=-m_{13})_{(10)}^{-23+33}
=\\[4mm] & &~~~~
-\left({[l_{13} - l_{11}] [l_{13} + l_{33} + 3]
[l_{13} + l_{31} + 2]
\over [l_{13} - l_{23}]}
\right)^{1/2}
(m;~m_{43}=-m_{13})_{(00)}^{-31}
,\\[6mm]
& &E_{23}(m;~m_{43}=-m_{13})_{(00)}=0.\\
& &~~~~~~~~ ~~~ ~~~ ~~~ ~~ ~~~ ~~~ ~~ ~
{}~~~ ~~ ~~~ ~~~ ~~~ ~~~~ ~~~ ~~ ~~ ~~ ~~ ~~
{}~~ ~~ ~~ ~~ ~~~~ ~~ ~~ ~~ ~~
{}~ ~~ ~~ ~~ ~~ ~~ ~~ ~~ (3.27)\\
\end{eqnarray*}
 It is not difficult to show that at the
classical limit when $q=1$ the matrix elements
(3.26) and (3.27) take respectively the values of
the matrix elements
of the $gl(2/2)$-generators $e_{32}$ and $e_{23}$
given in {\bf I$^{\ast}$} (from (I$^{\ast}$.82) to
(I$^{\ast}$.97)).\\

   As far as Propositions 8 and 9 in {\bf I$^{\ast}$}
(as well as all
the next Propositions there) are concerned
they can also be extended to the present case of
the q-deformations,
namely, if we put ~$m_{13}=m_{23}$~ and ~$m_{33}>-m_{13}$~
or ~$m_{13}>m_{23}$~ and ~$m_{33}=-m_{13}$~
the class 1 indecomposible module
{}~$W^{q}([m;~m_{43}=-m_{13}])$~ is decomposed
into irreducible ~$U_{q}[gl(2) \oplus gl(2)]$~ modules
in the same way as its classical analogue
{}~$W([m;~m_{43}=-m_{13}])$~ is decomposed into
irreducible ~$gl(2) \oplus gl(2)$~ modules in
(I$^{\ast}$.99)  or (I$^{\ast}$.103), respectively. Therefore,
these decompositions and the decompositions schemes for
all the remaining indecomposible and nontypical modules
are not necessary to be exposed again here unlike the
matrix elements which, although have similar structures
with their classical analogues, are not the same.
In order to obtain the transformations of a nontypical
module of any class we simply apply the procedure
(3.A.1)--(3.A.3) to the corresponding case. The direct
computations show that they can be
obtained immediately
from their classical analogues in {\bf I$^{\ast}$}
replacing the brackets "(~~~)" by the notation for
the quantum deformation  "[~~~]" and then removing
the modulus after using the rule (2.22), i.e.,
the rule (I.4.44).
As examples, we gave
explicitly only the matrix elements for the class 1
nontypical
representation (see (3.26) and (3.27)).
{}From now on, the matrix elements for the nontypical
representations of all the  next classes will be placed
in the Appendix , since they are too cumbersome.
\\[2mm]

{\bf B. The class 2 nontypical representations}\\

   This class corresponds to the case (3.2), namely,
\begin{tabbing}
\=
{}~~~~~~~~~~~~~~~~~~ \= class 1~~~
\= $l_{13} + l_{43} + 3$ \= = 0,
{}~~ \= $\Leftrightarrow$~~~ \= $m_{13} + m_{43} - 1$ = 0
{}~~~~~~~~~~~~\=\kill
\>   \> class 2   \>$l_{13} + l_{43} + 3$ \> $\neq 0$
\> $\Leftrightarrow$ \> $m_{13} + m_{43}$  $\neq 0$,\\[2mm]
\>   \>           \>$l_{23} + l_{33} + 3$ \> = 0
\> $\Leftrightarrow$ \> $m_{23} + m_{33}$  = 0;
\> (3.2)
\end{tabbing}
The signature of the induced modules is
{}~$[m;~m_{33}=-m_{23}]
:=[m_{13}, m_{23}, -m_{23}, m_{43}]$~ :
$$W([m;~m_{33}=-m_{23}])
=W([m_{13}, m_{23}, -m_{23}, m_{43}])
\eqno(3.28)$$

   {\it Proposition} 5: The vectors
\begin{eqnarray*}
{}~~~~~~~~~~~~\stackrel{{\small inv}}
{(m_{11}, m_{31})}_{(2)}
& = &
{[2] \over 2}\left\{
\left({[l_{13} - l_{23} - 1]}\over
{[l_{13} - l_{23} + 1]} \right)^{1/2}
(m;~m_{33}=-m_{23})_{(11)}^{-13-23+33+43}
\right. \\[4mm]
& & \left.
+\left({[l_{23} + l_{43} + 4]}\over
{[l_{23} + l_{43} + 2]} \right)^{1/2}
(m;~m_{33}=-m_{23})_{(20)}^{-13-23+33+43}\right\}
\\& &~~~~~~~~~~~~~~~~~~~~~~~~~~~~~~
{}~~ ~ ~ ~~  ~ ~ ~~ ~ ~ ~ ~ ~ ~ ~ ~ ~ ~ ~ ~ ~
{}~~ ~ ~ ~~ ~~~ ~~ ~(3.29)
\end{eqnarray*}
and\\[2mm]
\begin{eqnarray*}
{}~~~~~~~~~~~~(m_{11}, m_{31})_{(2)} & := &
(m;~m_{33}=-m_{23})_{(2)}^{-13-23+33+43}\\[4mm]
& = &
{[2] \over 2}\left\{
\left({[l_{13} - l_{23} - 1]}\over
{[l_{13} - l_{23} + 1]} \right)^{1/2}
(m;~m_{33}=-m_{23})_{(11)}^{-13-23+33+43}
\right. \\[4mm]
& & \left.
-\left({[l_{23} + l_{43} + 4]}\over
{[l_{23} + l_{43} + 2]} \right)^{1/2}
(m;~m_{33}=-m_{23})_{(20)}^{-13-23+33+43}\right\}
\\& &~~~~~~~~~~~~~~~~~~~~~~~~~~~~~~
{}~~ ~ ~ ~~  ~ ~ ~~ ~ ~ ~ ~ ~ ~ ~ ~ ~ ~ ~ ~ ~
{}~~ ~ ~ ~~ ~~~ ~~ ~(3.30)
\end{eqnarray*}

form respectively linear spaces
\begin{eqnarray*}
{}~~~~~~~~\stackrel{{\small inv}}{V}^{~q}_{(2)} & := &
\stackrel{{\small inv}}{V}^{~q}_{(2)}
([m\pm 1;~m_{33}=-m_{23}])\\
& =  & {\normalsize lin. env.}\{
\stackrel{{\small inv}}{(m_{11}, m_{31})}_{(2)} \|
{}~ m_{13} - m_{11} - 1, m_{11} - m_{23} + 1,\\
& &
-m_{23} - m_{31} + 1, m_{31} - m_{43} - 1 \in
{\bf Z}_{+}\}
{}~~~~~~~~~~~~~~~~~~~~~~~~~~(3.31)
\end{eqnarray*}
and
\begin{eqnarray*}
{}~~~~~~~~V^{q}_{(2)} & := &
V^{q}_{(2)}([m\pm 1;~m_{33}=-m_{23}])\\
& =  &{\normalsize lin. env.}\{(m_{11}, m_{31})_{(2)} \|
{}~m_{13} - m_{11} - 1, m_{11} - m_{23} + 1,\\
& &
-m_{23} - m_{31} + 1, m_{31} - m_{43} - 1 \in
{\bf Z}_{+}\},
{}~~~~~~~~~~~~~~~~~~~~~~~~~~(3.32)
\end{eqnarray*}
which being irreducible
{}~$U_{q}[gl(2) \oplus gl(2)]$~ modules
satisfy
$$\stackrel{{\small inv}}
{V}^{~q}_{(2)}([m\pm 1;~m_{33}=-m_{23}])
=~~\stackrel{{\small inv}}
{\frame{\shortstack{
{\small ~-1,-1,1,1}}~}}_{~(2)},
\eqno(3.33)$$
i.e.,
$$\stackrel{{\small inv}}
{V}^{~q}_{(2)}([m\pm 1;~m_{33}=-m_{23}])
{}~\subset I^{q}_{2}$$
and
$$V^{q}_{(2)}([m\pm 1;~m_{33}=-m_{23}])
=~~\stackrel{{\small noninv}}
{\frame{\shortstack{
{\small ~-1,-1,1,1}}~}}_{~(2)},
\eqno(3.34)$$
i.e.,
$$V^{q}_{(2)}([m\pm 1;~m_{33}=-m_{23}])
{}~\cap I^{q}_{2} =\O.$$
\vspace*{2mm}

 {\bf C. The class 3 nontypical representations}\\

   This class corresponds to the case (3.3)
\begin{tabbing}
\=
{}~~~~~~~~~~~~~~~~~~ \= class 1~~~
\= $l_{13} + l_{43} + 3$ \= = 0,
{}~~ \= $\Leftrightarrow$~~~ \= $m_{13} + m_{43} - 1$ = 0
{}~~~~~~~~~~~~\=\kill
\>   \> class 3   \>$l_{23} + l_{43} + 3$ \> = 0
\> $\Leftrightarrow$ \> $m_{23} + m_{43}  - 1$  = 0;
\> (3.3)
\end{tabbing}
and the signature
{}~$[m;~m_{43}=-m_{23}+1]
:=[m_{13}, m_{23}, m_{33}, -m_{23}+1]$~ :
$$W([m;~m_{43}=-m_{23}+1])
=W([m_{13}, m_{23}, m_{33}, -m_{23}+1])
\eqno(3.35)$$

   {\it Proposition} 6: The vectors
\begin{eqnarray*}
{}~~~~~~~~~~~~\stackrel{{\small inv}}
{(m_{11}, m_{31})}_{(3)}
& = & {[2] \over 2}\left\{
\left({[l_{13} - l_{23} - 1]}\over
{[l_{13} - l_{23} + 1]} \right)^{1/2}
(m;~m_{43}=-m_{23}+1)_{(11)}^{-13-23+33+43}
\right. \\[4mm]
& & \left.
- \left({[l_{23} + l_{33} + 4]}\over
{[l_{23} + l_{33} + 2]} \right)^{1/2}
(m;~m_{43}=-m_{23}+1)_{(20)}^{-13-23+33+43}\right\}
\\& &~~~~~~~~~~~~~~~~~~~~~~~~~~~~~~
{}~~ ~ ~ ~~  ~ ~ ~~ ~ ~ ~ ~ ~ ~ ~ ~ ~ ~ ~ ~ ~
{}~~ ~ ~ ~~ ~~~ ~~ ~~(3.36)
\end{eqnarray*}
and\\[2mm]
\begin{eqnarray*}
{}~~~~~~~~~~~~(m_{11}, m_{31})_{(3)} & := &
(m;~m_{43}=-m_{23}+1)_{(3)}^{-13-23+33+43}\\[4mm]
& = & {[2] \over 2}\left\{
\left({[l_{13} - l_{23} - 1]}\over
{[l_{13} - l_{23} + 1]} \right)^{1/2}
(m;~m_{43}=-m_{23}+1)_{(11)}^{-13-23+33+43}
\right. \\[4mm]
& & \left.
+ \left({[l_{23} + l_{33} + 4]}\over
{[l_{23} + l_{33} + 2]} \right)^{1/2}
(m;~m_{43}=-m_{23}+1)_{(20)}^{-13-23+33+43}\right\}
\\& &~~~~~~~~~~~~~~~~~~~~~~~~~~~~~~
{}~~ ~ ~ ~~  ~ ~ ~~ ~ ~ ~ ~ ~ ~ ~ ~ ~ ~ ~ ~ ~
{}~~ ~ ~ ~~ ~~~ ~~ ~~(3.37)
\end{eqnarray*}

form respectively linear spaces
\begin{eqnarray*}
{}~~~~~~~~\stackrel{{\small inv}}{V}^{~q}_{(3)} & := &
\stackrel{{\small inv}}{V}^{~q}_{(3)}
([m\pm 1;~m_{43}=-m_{23}+1])\\
& =  & {\normalsize lin. env.}\{
\stackrel{{\small inv}}{(m_{11}, m_{31})}_{(3)} \|
{}~ m_{13} - m_{11} - 1, m_{11} - m_{23} + 1,\\
& &
m_{33} - m_{31} + 1, m_{31} + m_{23} - 2 \in
{\bf Z}_{+}\}
{}~~~~~~~~~~~~~~~~~~~~~~~~~~~~~~~~~(3.38)
\end{eqnarray*}
and
\begin{eqnarray*}
{}~~~~~~~~V^{q}_{(3)} & := &
V^{q}_{(3)}([m\pm 1;~m_{43}=-m_{23}+1])\\
& =  &{\normalsize lin. env.}\{(m_{11}, m_{31})_{(3)} \|
{}~m_{13} - m_{11} - 1, m_{11} - m_{23} + 1,\\
& &
m_{33} - m_{31} + 1, m_{31} + m_{23} - 2\in
{\bf Z}_{+}\},
{}~~~~~~~~~~~~~~~~~~~~~~~~~~~~~~~(3.39)
\end{eqnarray*}
which being irreducible
{}~$U_{q}[gl(2) \oplus gl(2)]$~ modules
satisfy
$$\stackrel{{\small inv}}
{V}^{~q}_{(3)}([m\pm 1;~m_{43}=-m_{23}+1])
=~~\stackrel{{\small inv}}
{\frame{\shortstack{
{\small ~-1,-1,1,1}}~}}_{~(3)},
\eqno(3.40)$$
i.e.,
$$\stackrel{{\small inv}}
{V}^{~q}_{(3)}([m\pm 1;~m_{43}=-m_{23}+1])
{}~\subset I^{q}_{3}$$
and
$$V^{q}_{(3)}([m\pm 1;~m_{43}=-m_{23}+1])
=~~\stackrel{{\small noninv}}
{\frame{\shortstack{
{\small ~-1,-1,1,1}}~}}_{~(3)},
\eqno(3.41)$$
i.e.,
$$V^{q}_{(3)}([m\pm 1;~m_{43}=-m_{23}+1])
{}~\cap I^{q}_{3} =\O.$$
\vspace*{2mm}

 {\bf D. The class 4 nontypical representations}\\

   This class corresponds to the case (3.4)
\begin{tabbing}
\=
{}~~~~~~~~~~~~~~~~~~ \= class 1~~~
\= $l_{13} + l_{43} + 3$ \= = 0,
{}~~ \= $\Leftrightarrow$~~~ \= $m_{13} + m_{43} - 1$ = 0
{}~~~~~~~~~~~~\=\kill
\>   \> class 4   \>$l_{13} + l_{33} + 3$ \> = 0
\> $\Leftrightarrow$ \> $m_{13} + m_{33} + 1$  = 0;
\> (3.4)\\[4mm]
\end{tabbing}
and the signature
{}~$[m;~m_{33}=-m_{13}-1]
:=[m_{13}, m_{23}, -m_{13}-1, m_{43}]$~ :
$$W([m;~m_{13}=-m_{13}-1])
=W([m_{13}, m_{23}, -m_{13}-1, m_{43}])
\eqno(3.42)$$

   {\it Proposition} 7: The vectors
\begin{eqnarray*}
{}~~~~~~~~~~~~\stackrel{{\small inv}}
{(m_{11}, m_{31})}_{(4)}
& = & {[2] \over 2}\left\{
- \left({[l_{13} - l_{23} + 1]}\over
{[l_{13} - l_{23} - 1]} \right)^{1/2}
(m;~m_{33} = -m_{13}-1)_{(11)}^{-13-23+33+43}
\right. \\[4mm]
& & \left.
+ \left({[l_{13} + l_{43} + 4]}\over
{[l_{13} + l_{43} + 2]} \right)^{1/2}
(m;~m_{33}=-m_{13}-1)_{(20)}^{-13-23+33+43}\right\}
\\& &~~~~~~~~~~~~~~~~~~~~~~~~~~~~~~
{}~~ ~ ~ ~~  ~ ~ ~~ ~ ~ ~ ~ ~ ~ ~ ~ ~ ~ ~ ~ ~
{}~~ ~ ~ ~~ ~~~ ~~ ~(3.43)
\end{eqnarray*}
and\\[2mm]
\begin{eqnarray*}
{}~~~~~~~~~~~~(m_{11}, m_{31})_{(4)} & := &
(m;~m_{33}=-m_{13}-1)_{(4)}^{-13-23+33+43}\\[4mm]
& = & {[2] \over 2}\left\{
-\left({[l_{13} - l_{23} + 1]}\over
{[l_{13} - l_{23} - 1]} \right)^{1/2}
(m;~m_{33}=-m_{13}-1)_{(11)}^{-13-23+33+43}
\right.\\[4mm]
& & \left.
- \left({[l_{13} + l_{43} + 4]}\over
{[l_{13} + l_{43} + 2]} \right)^{1/2}
(m;~m_{33}=-m_{13}-1)_{(20)}^{-13-23+33+43}\right\}
\\& &~~~~~~~~~~~~~~~~~~~~~~~~~~~~~~
{}~~ ~ ~ ~~  ~ ~ ~~ ~ ~ ~ ~ ~ ~ ~ ~ ~ ~ ~ ~ ~
{}~~ ~ ~ ~~ ~~~ ~~ ~(3.44)
\end{eqnarray*}

form respectively linear spaces
\begin{eqnarray*}
{}~~~~~~~~\stackrel{{\small inv}}{V}^{~q}_{(4)} & := &
\stackrel{{\small inv}}{V}^{~q}_{(4)}
([m\pm 1;~m_{33}=-m_{13}-1])\\
& =  & {\normalsize lin. env.}\{
\stackrel{{\small inv}}{(m_{11}, m_{31})}_{(4)} \|
{}~ m_{13} - m_{11} - 1, m_{11} - m_{23} + 1,\\
& &
-m_{13} - m_{31}, m_{31} - m_{43} - 1 \in {\bf Z}_{+}\}
{}~~~~~~~~~~~~~~~~~~~~~~~~~~~~~~~~~(3.45)
\end{eqnarray*}
and
\begin{eqnarray*}
{}~~~~~~~~V^{q}_{(4)} & := &
V^{q}_{(4)}([m\pm 1;~m_{33}=-m_{13}-1])\\
& =  &{\normalsize lin. env.}\{(m_{11}, m_{31})_{(4)} \|
{}~m_{13} - m_{11} - 1, m_{11} - m_{23} + 1,\\
& &
-m_{13} - m_{31}, m_{31} - m_{43} - 1 \in
{\bf Z}_{+}\},
{}~~~~~~~~~~~~~~~~~~~~~~~~~~~~~~~~(3.46)
\end{eqnarray*}
which being irreducible
{}~$U_{q}[gl(2) \oplus gl(2)]$~ modules
satisfy
$$\stackrel{{\small inv}}
{V}^{~q}_{(4)}([m\pm 1;~m_{33}=-m_{13}-1])
=~~\stackrel{{\small inv}}
{\frame{\shortstack{
{\small ~-1,-1,1,1}}~}}_{~(4)},
\eqno(3.47)$$
i.e.,
$$\stackrel{{\small inv}}
{V}^{~q}_{(4)}([m\pm 1;~m_{33}=-m_{13}-1])
{}~\subset I^{q}_{4}$$
and
$$V^{q}_{(4)}([m\pm 1;~m_{33}=-m_{13}-1])
=~~\stackrel{{\small noninv}}
{\frame{\shortstack{
{\small ~-1,-1,1,1}}~}}_{~(4)},
\eqno(3.48)$$
i.e.,
$$V^{q}_{(4)}([m\pm 1;~m_{33}=-m_{13}-1])
{}~\cap I^{q}_{4} =\O.$$
\vspace*{2mm}

 {\bf E. The class 5 nontypical representations}\\

   This class corresponds to the case (3.5)
\begin{tabbing}
\=
{}~~~~~~~~~~~~~~~~~~ \= class 5~~~
\= $l_{13} + l_{43} + 3$ \= = 0,
{}~~ \= $\Leftrightarrow$~~~ \= $m_{13} + m_{43} - 1$ = 0
{}~~~~~~~~~~~~\=\kill
\>   \> class 5   \>$l_{13} + l_{43} + 3$ \> = 0
\> $\Leftrightarrow$ \> $m_{13} + m_{43}$  = 0,\\[2mm]
\>   \>           \>$l_{23} + l_{33} + 3$ \> = 0
\> $\Leftrightarrow$ \> $m_{23} + m_{33}$  = 0.
\> (3.5)
\end{tabbing}
and the signature
{}~$[m;~m_{33}=-m_{23},m_{43}=-m_{13}]
:=[m_{13}, m_{23}, -m_{23}, -m_{13}]$~ :
$$W([m;~m_{33}=-m_{23},m_{43}=-m_{13}])
=W([m_{13}, m_{23}, -m_{23}, -m_{13}])
\eqno(3.49)$$

   {\it Proposition} 8: The vectors
\begin{eqnarray*}
{}~~~~~~~~~~~~\stackrel{{\small inv}}
{(m_{11}, m_{31})}_{(5)}
& = & {[2] \over 2}
\left({[l_{13} - l_{23} - 1]}\over
{[l_{13} - l_{23} + 1]} \right)^{1/2}\\[4mm]
& & \times
\left\{
(m;~m_{33}=-m_{23},m_{43}=-m_{13})_{(11)}^{-13-23+33+43}
\right.\\[4mm]
& &
\left.
+ (m;~m_{33}=-m_{23},m_{43}=-m_{13})_{(20)}^{-13-23+33+43}
\right\}
\\& &~~~~~~~~~~~~~~~~~~~~~~~~~~~~~~
{}~~ ~ ~ ~~  ~ ~ ~~ ~ ~ ~ ~ ~ ~ ~ ~ ~ ~ ~ ~ ~
{}~~ ~ ~ ~~ ~~~ ~~ ~(3.50)
\end{eqnarray*}
and\\[2mm]
\begin{eqnarray*}
{}~~~~~~~~~~~~(m_{11}, m_{31})_{(5)} & := &
(m;~m_{33}=-m_{23},m_{43}=-m_{13})_{(5)}^{-13-23+33+43}\\[4mm]
& = & {[2] \over 2}
\left({[l_{13} - l_{23} - 1]}\over
{[l_{13} - l_{23} + 1]} \right)^{1/2}\\[4mm]
& & \times
\left\{
(m;~m_{33}=-m_{23},m_{43}=-m_{13})_{(11)}^{-13-23+33+43}
\right.\\[4mm]
& &
\left.
- (m;~m_{33}=-m_{23},m_{43}=-m_{13})_{(20)}^{-13-23+33+43}
\right\}
\\& &~~~~~~~~~~~~~~~~~~~~~~~~~~~~~~
{}~~ ~ ~ ~~  ~ ~ ~~ ~ ~ ~ ~ ~ ~ ~ ~ ~ ~ ~ ~ ~
{}~~ ~ ~ ~~ ~~~ ~~ (3.51)
\end{eqnarray*}
form respectively linear spaces
\begin{eqnarray*}
{}~~~~~~~~\stackrel{{\small inv}}{V}^{~q}_{(5)} & := &
\stackrel{{\small inv}}{V}^{~q}_{(5)}
([m\pm 1;~m_{33}=-m_{23},m_{43}=-m_{13}])\\
& =  & {\normalsize lin. env.}\{
\stackrel{{\small inv}}{(m_{11}, m_{31})}_{(5)} \|
{}~ m_{13} - m_{11} - 1, m_{11} - m_{23} + 1,\\
& &
-m_{23} - m_{31} + 1, m_{31} + m_{13} - 1 \in {\bf Z}_{+}\}
{}~~~~~~~~~~~~~~~~~~~~~~~~~~~~~~~~~~(3.52)
\end{eqnarray*}
and
\begin{eqnarray*}
{}~~~~~~~~V^{q}_{(5)} & := &
V^{q}_{(5)}([m\pm 1;~m_{33}=-m_{23},m_{43}=-m_{13}])\\
& =  &{\normalsize lin. env.}\{(m_{11}, m_{31})_{(5)} \|
{}~m_{13} - m_{11} - 1, m_{11} - m_{23} + 1,\\
& &
-m_{23} - m_{31} + 1, m_{31} + m_{13} - 1 \in
{\bf Z}_{+}\},
{}~~~~~~~~~~~~~~~~~~~~~~~~~~~~~~~~~(3.53)
\end{eqnarray*}
which being irreducible ~$U_{q}[gl(2) \oplus gl(2)]$~ modules
satisfy
$$\stackrel{{\small inv}}
{V}^{~q}_{(5)}([m\pm 1;~m_{33}=-m_{23},m_{43}=-m_{13}])
=~~\stackrel{{\small inv}}
{\frame{\shortstack{
{\small ~-1,-1,1,1}}~}}_{~(5)},
\eqno(3.54)$$
i.e.,
$$\stackrel{{\small inv}}
{V}^{~q}_{(5)}([m\pm 1;~m_{33}=-m_{23},m_{43}=-m_{13}])
{}~\subset I^{q}_{5}$$
and
$$V^{q}_{(5)}([m\pm 1;~m_{33}=-m_{23},m_{43}=-m_{13}])
=~~\stackrel{{\small noninv}}
{\frame{\shortstack{
{\small ~-1,-1,1,1}}~}}_{~(5)},
\eqno(3.55)$$
i.e.,
$$V^{q}_{(5)}([m\pm 1;~m_{33}=-m_{23},m_{43}=-m_{13}])
{}~\cap I^{q}_{5} =\O.$$
\vspace*{2mm}

  The decomposition of
{}~$W^{q}([m;~m_{33} = -m_{23},m_{43} = -m_{13}])$~
is the same as given in (I$^{\ast}$.227). Unlike the maximal
invariant
subspaces ~$I^{q}_{k}$~,
$k=1,2,3,4$~, of the previous classes the maximal invariant
subspace ~$I^{q}_{5}$~ is indecomposible and decomposed into
three invariant subspaces ~$I^{q,0}_{5}$~, ~$I^{q,1}_{5}$~,
{}~$I^{q,2}_{5}$~, where ~$I^{q,0}_{5}$~ is an irreducible
and nontypical module with a signature
$$[m\pm 1;~m_{33}=-m_{23},m_{43}=-m_{13}]\eqno(3.56)$$
while every of ~$I^{q,1}_{5}$~ and ~$I^{q,2}_{5}$~
is indecomposible and contains ~$I^{q,0}_{5}$~
as a maximal invariant subspace. The factor spaces
{}~$I^{q,1}_{5}/I^{q,0}_{5}$ and ~$I^{q,2}_{5}/I^{q,0}_{5}$~
in turn carry nontypical representations of
$U_{q}[gl(2/2)]$
with signatures
$$[m;~m_{33} = -m_{23},m_{43} = -m_{13}]^{-23+33}\eqno(3.57)$$
and
$$[m;~m_{33} = -m_{23},m_{43} = -m_{13}]^{-13+43},\eqno(3.58)$$
respectively.\\[5mm]

\begin{flushleft}
{\bf IV. FINITE--DIMENSIONAL REPRESENTATIONS OF
{}~U$_{q}$[gl(2/2)]}
\end{flushleft}
\vspace*{2mm}

   As in {\bf I$^{\ast}$} we denote by ~$\Im$~ the class of
finite--dimensional irreducible ~$U_{q}[gl(2/2)]$~ modules
{}~$W^{q}([m_{13},m_{23},m_{33},m_{43}])$~ determined in
{\bf I} and in this paper. The modules ~$W^{q}\in \Im$~
are characterized by the signatures (1.6) which represent
ordered sets of all possible complex numbers
$m_{13},~m_{23},~m_{33}$ and $m_{43}$ satisfying the
conditions (1.3)
$$m_{13} - m_{23},~  m_{33}- m_{43} \in {\bf Z}_{+}.
\eqno(4.1)$$
The typical modules are those for which
(1.9) holds, while the indecomposible modules carrying
nontypical representations of ~$U_{q}[gl(2/2)]$~, are
those for which one of the conditions (3.1)-(3.5) is
fulfilled. The transformations of all the nontypical
modules under the actions of ~$U_{q}[gl(2/2)]$~ are
completely defined through the actions (I.4.43) of the
even generators and the actions of the odd Chevalley
generators $E_{23}$ and $E_{32}$ written down in (3.26)
and (3.27) for the class 1 and in the Appendix for the
other classes. The transformations of all the typical
modules have already given in {\bf I} by the formulae
(I.4.43)-(I.4.46).\\

   {\it Proposition} 9: If ~$W^{q}$~ is a
{}~$U_{q}[gl(2/2)]$~ fidirmod, then ~$W^{q}\in \Im$.\\

   We shall sketch the proof.
Let us denote by $\{x_{\Lambda_{k}}\}$ the set of all
the eigenvectors $x_{\Lambda_{k}}$, $k\in {\bf Z_{+}}$,
of the Cartan
generators ~$E_{ii}$, $i=1,2,3,4$, such that they
are annihilated by the generators $E_{12}$ and $E_{34}$.
Acting on every $x_{\Lambda_{k}} \in \{x_{\Lambda_{k}}\}$
by all possible
elements of ~$U_{q}^{0}:=U_{q}[gl(2)\oplus gl(2)]$~
(cf. (I.4.15)) we obtain a set  $\{x^{k}\}$
$$\{x^{k}\}=U_{q}^{0}x_{\Lambda_{k}}\eqno(4.2)$$
which is a finite--dimensional irreducible
{}~$U_{q}[gl(2)\oplus gl(2)]$ module and
spanned on all basis vectors $x^{k}$ determined up
to multiplicative constants in the forms
$$x^{k}=
c(x^{k})(E_{21})^{n_{1}}(E_{43})^{n_{2}}x_{\Lambda_{k}}~;~~
n_{1},~n_{2} \in {\bf Z_{+}}.
\eqno(4.3)$$
Since $q$ is generic (i.e., there are not cyclic
representations) and  ~$W^{q}$~ is a
finite--dimensional irreducible ~$U_{q}[gl(2/2)]$ module
it can be shown that the union $\{x\}$ of all the sets
($U_{q}[gl(2/2)]$ fidirmods) $\{x^{k}\}$ cover ~$W^{q}$
whole and every of the odd generators (e.g., $E_{23}$)
intertwining the sets $\{x^{k}\}$ has to vanish
in some $\{x^{0}\}$. Hence, there exists always in
{}~$W^{q}$~ an eigenvector ~$x_{\Lambda}^{0}$~ of the
Cartan generators ~$E_{ii}$, $i=1,2,3,4$,
$$E_{ii}x_{\Lambda}^{0}=m_{i3}x_{\Lambda}^{0}
\eqno(4.4)$$
such that it is annihilated  by
the generators $E_{12}$ and $E_{34}$
$$E_{12}x_{\Lambda}^{0}=E_{34}x_{\Lambda}^{0}=0
\eqno(4.5)$$
and simultaneously by the generator $E_{23}$
$$E_{23}x_{\Lambda}^{0}=0\eqno(4.6)$$
Therefore, ~$x_{\Lambda}^{0}$~ is a highest weight vector
of the highest weight $[m]=[m_{13},m_{23},m_{33},m_{43}]$
with respect to  both ~$U_{q}[gl(2)\oplus gl(2)]$~ and
{}~$U_{q}[gl(2/2)]$. In other words, the
{}~$U_{q}[gl(2)\oplus gl(2)]$ fidirmod
$$V_{(00)}^{q}:=\{x^{0}\}=U_{q}^{0}x_{\Lambda}^{0}
\eqno(4.7)$$
and the $U_{q}[gl(2/2)]$ fidirmod ~$W^{q}$~ have one and
the same signature $[m]$. The module $V_{(00)}^{q}$,
however, is a tensor product
$$V_{(00)}^{q}=V_{(00),l}^{q} \oplus V_{(00),r}^{q}
\eqno(4.8)$$
between a $U_{q}[gl(2)_{l}]$ fidirmod $V_{(00),l}^{q}$~
and a $U_{q}[gl(2)_{r}]$ fidirmod $V_{(00),r}^{q}$~
which in turn are labeled by the signatures
$[m]_{l}=[m_{13},m_{23}]$ and $[m]_{r}=[m_{33},m_{43}]$,
respectively. As is well known that finite--dimensional
representations of $U_{q}[gl(2)]$ are q-deformations of
the finite--dimensional representations of the classical
$gl(2)$ and the GZ bases are invariant under the
q-deformations. Therefore, the classical conditions
$$
m_{13}-m_{23} \in {\bf Z_{+}},~~
m_{33}- m_{43}\in {\bf Z_{+}}$$
imposed on the complex numbers
$m_{13}$, $m_{23}$, $m_{33}$, $m_{43}$ are still valid
in the present case and coincide with (4.1),
i.e., ~$W^{q} \in \Im$.\\

     We can conclude
that a ~$U_{q}[gl(2/2)]$ fidirmod ~$W^{q}$~ is also a
{}~$U_{q}[sl(2/2)]$~ fidirmod and vice versa.
The quantum superalgebra ~$U_{q}[sl(2/2)]$~
is generated completely by
the Cartan generators
$$h_{1}=E_{11}-E_{22},~h_{2}=E_{22}+E_{33},
{}~h_{3}=E_{33}-E_{44}\eqno(4.9)$$
defined in (2.1) (or (I.3.1)) and by the other
{}~$U_{q}[gl(2/2)]$-Chevalley generators ~$E_{ij}$,~
$i\neq j=1,2,3,4$, satisfying the defining
relations (2.1)-(2.4) (or (I.3.1)-(I.3.4). A signature
(highest weight) of a ~$U_{q}[sl(2/2)]$ fidirmod is
defined again as an ordered set of eigenvalues of the
Cartan generators $h_{i}$, $i=1,2,3$, on a highest
weight vector. Since ~$U_{q}[gl(2/2)]$~ and
{}~$U_{q}[sl(2/2)]$~ have ones and the same Chevalley
(nondiagonal) generators a ~$U_{q}[gl(2/2)]$-highest
weight vector is also a
{}~$U_{q}[sl(2/2)]$-highest weight vector and vice versa.
The ~$U_{q}[gl(2/2)]$ fidirmod
{}~$W^{q}[m_{13},m_{23},m_{33},m_{43}]$~ considered as a
{}~$U_{q}[sl(2/2)]$ fidirmod is labeled by the numbers
$$\alpha_{1}=m_{13}-m_{23},~~
\alpha_{2}=m_{23}+m_{33},~~
\alpha_{3}=m_{33}-m_{43},\eqno(4.10)$$
where ~$\alpha_{i}$, $i=1,2,3$, are eigenvalues of $h_{i}$.
Therefore, any ~$U_{q}[sl(2/2)]$-signature (i.e., the
signature of a module ~$W^{q}$~ considered as a
{}~$U_{q}[sl(2/2)]$ fidirmod) is uniquely determined from
some  ~$U_{q}[gl(2/2)]$-signature.
If $m_{13}$, $m_{23}$, $m_{33}$, $m_{43}$
labeling ~$U_{q}[gl(2/2)]$ fidirmods take all values
consistent with (4.1) then the triple
$(\alpha_{1}, \alpha_{2}, \alpha_{3})$ runs over all
labels (signatures) for the ~$U_{q}[sl(2/2)]$ fidirmods.
Vice versa, a ~$U_{q}[sl(2/2)]$ fidirmod can be extended
to (inequivalent copies of) ~$U_{q}[gl(2/2)]$ fidirmods
with signatures determined from the
{}~$U_{q}[sl(2/2)]$-signature and by setting
$E_{44}x_{\Lambda}^{0}= \alpha x_{\Lambda}^{0}$,
(~$\alpha \equiv m_{43}$), on the highest weight vector
{}~$x_{\Lambda}^{0}$. We have shown that the following
Proposition holds\\

  {\it Proposition} 10: The class ~$\Im$~ contains all
finite-dimensional irreducible  ~$U_{q}[sl(2/2)]$ modules.\\

   The next step one can make is to consider finite-dimensional
irreducible representations of the quantum superalgebra
{}~$U_{q}[A(1/1)]$.\\[5mm]

\begin{flushleft}
{\bf V. CONCLUSION}
\end{flushleft}
\vspace*{2mm}

    We have completed our programme on an explicit
construction at generic deformation parameter $q$
of all (typical and nontypical) finite--dimensional
representations of the
quantum Lie superalgebra $U_{q}[gl(2/2)]$.
The construction method proposed in {\bf I}
allowed us to construct the induced modules of
(certainly, not only) $U_{q}[gl(2/2)]$ with
basis systems convenient for finding all possible
finite-dimensional irreducible modules and
representations of this quantum superalgebra.
The previous paper {\bf I} is devoted to the general
construction procedure with an accent on the typical
representations, while in this paper the indecomposible
representations were considered and classified.
Here, the nontypical representations of all classes as
irreducible representations extracted from the
indecomposible ones were constructed explicitly.
As we can see, all the results obtained in {\bf I}
and in the present paper coincide at $q=1$ with
the classical ones in Ref. 14 and 15.
It turns out that
although the quantum
deformation gives rise to some specific difficulties
and makes the present construction more cumbersome the
resemblance between the quantum structures and
the non-deformation ones $^{14,15}$ is remarkable.
As far as the case of
$q$ being roots of unity is concerned it is a subject of a
later publication. In the latter case, however,
the structures of $U_{q}[gl(2/2)]$-modules are drastically
different in comparison with the structures of
$gl(2/2)$-modules $^{14,15}$.\\

  Putting the results of Refs. 1,14,15 and the
present paper all together we arrive at the following
conclusion\\

   {\it Proposition} 11: The finite-dimensional representations
of the quantum superalgebra ~$U_{q}[gl(2/2)]$~ are quantum
deformations (q-deformations) of the finite-dimensional
representations of the superalgebra ~$gl(2/2)$.\\

  Certainly, our construction procedure proposed in {\bf I}
and the present paper for ~$U_{q}[gl(2/2)]$~ is applicable
to higher rank ~$U_{q}[gl(m/n)]$~ and other quantum (super)
groups, for which Proposition 11 may remain valid. Then our
approach may have some advantage as it is worthy to mention
that the theory of representations and especially of the
nontypical ones is far from being complete even for the
nondeformed superalgebras. In particular, the dimensions
of the nontypical representations are unknown unless the
ones for $sl(1/n)$ computed recently in Ref. 19.  Based
on the generalizations of the concept of the GZ basis
(see, for example, Refs. 20 and 21)
the matrix elements of all nontypical
representations were
computed only for $sl(1/n)$ and
$gl(1/n)$ (see Refs. 22).
Later, the essentially typical
representations of $gl(m/n)$ were also constructed $^{23}$.
So far, however, the GZ basis concept was not not defined
and presumably can not be defined for nontypical
$gl(m/n)$-modules with $m,n \geq 2$. This was the reason
to try to describe the nontypical modules in terms of the
basis of the even subalgebras. This approach was developed
so far for $gl(2/2)$ in Ref. 14 and for $U_{q}[gl(2/2)]$
in ${\bf I}$ and it turned appropriate for explicit
descriptions of all nontypical modules of $gl(2/2)$
(see Ref. 15) and $U_{q}[gl(2/2)]$, respectively. The approach
in ${\bf I}$, unlike some ealier approachs $^{14,20}$,
avoids, however, the use of the Clebsch--Gordan
coefficients which are not always known for higher rank
(quantum and classical) algebras.
Other extensions were made in Ref. 3 for all finite-dimensional
representations of $U_{q}[gl(1/n)]$  and in Ref. 6
for a class of finite-dimensional representations of
$U_{q}[gl(m/n)]$. To our best
knowledge, we give for the first time in ${\bf I}$ and the
present papers, respectively,
the explicit expressions for all typical
representations and all nontypical representations of
a quantum superalgebra $U_{q}[gl(m/n)]$ with $m,n\geq 2$.
\\[5mm]

\begin{flushleft}
{\bf ACKNOWLEDGMENTS}
\end{flushleft}
\vspace*{2mm}

   The authors are thankful to Professor Tch. Palev for
suggestions and
Professor I. Todorov for his interest in the present work.
N. A. K. would like to thank Professor Abdus Salam,
the International Atomic Energy Agency and UNESCO for
the kind hospitality at the High Energy Section of
the International Centre for Theoretical Physics,
Trieste, Italy.\\[5mm]
\begin{flushleft}
{\bf APPENDIX}:  Nontypical repersentation matrix elements of
$E_{32}$ and $E_{23}$
\end{flushleft}
\vspace*{2mm}

  {\bf A.1. Class 1}\\

   All the matrix elements of $E_{32}$ and $E_{23}$ for this
class nontypical representations have already given in (3.26)
and (3.27), respectively.\\[2mm]

  {\bf A.2. Class 2}\\

      The  matrix elements of $E_{32}$:

\begin{eqnarray*}
& &E_{32}(m;~m_{33}=-m_{23})_{(00)}=\\[4mm] & &~~~~
-\left({[l_{13} - l_{11}] [l_{31} - l_{43}] \over
[l_{13} - l_{23}] [l_{23} + l_{43} + 3]} \right)^{1/2}
(m;~m_{33}=-m_{23})_{(10)}^{-13+33+31}\\[4mm]
& &~~~~-\left({[l_{13} - l_{11}] [l_{23} + l_{31} + 3] \over
[l_{13} - l_{23}] [l_{23} + l_{43} + 3]} \right)^{1/2}
(m;~m_{33}=-m_{23})_{(10)}^{-13+31+43}\\[4mm]
& &~~~~-\left({[l_{11} - l_{23}] [l_{23} + l_{31}+3] \over
[l_{13} - l_{23}] [l_{23} + l_{43} + 3]} \right)^{1/2}
(m;~m_{33}=-m_{23})_{(10)}^{-23+43+31},\\[6mm]
& &E_{32}(m;~m_{33}=-m_{23})_{(10)}^{-13+33}=
\\[4mm] & &~~~~
-\left({[l_{13} - l_{11}-1] [l_{23} + l_{31} + 2] \over
[l_{13} - l_{23}-1] [l_{23} + l_{43} + 3]}
\right)^{1/2}
(m;~m_{33}=-m_{23})_{(11)}^{-13-13+33+43+31}\\[4mm]
& &~~~~-{1 \over (l_{13} - l_{23}-1)}
\left({[l_{13} - l_{23}]
[l_{11} - l_{23}]
[l_{23} + l_{31} + 2] \over
[2][l_{23} + l_{43} + 3]}
\right)^{1/2}\\[4mm]
& &~~~~\times
(m;~m_{33}=-m_{23})_{(2)}^{-13-23+33+43+31},\\[6mm]
& &E_{32}(m;~m_{33}=-m_{23})_{(10)}^{-13+43}=
\\[4mm] & &~~~~
\left({[l_{11} - l_{23}] [l_{23} + l_{31} + 3] \over
[l_{13} - l_{23}] [l_{23} + l_{43} + 4]}
\right)^{1/2}
(m;~m_{33}=-m_{23})_{(20)}^{-13-23+43+43+31}\\[4mm]
& &~~~~+\left({[l_{13} - l_{11}-1] [l_{31} - l_{43} - 1] \over
[l_{13} - l_{23} - 1] [l_{23} + l_{43} + 3]}
\right)^{1/2}
(m;~m_{33}=-m_{23})_{(11)}^{-13-13+33+43+31}\\[4mm]
& &~~~~+{[l_{13} + l_{43} + 3] \over
[l_{23} + l_{43} + 4][l_{13} - l_{23} - 1]}
\left({[l_{11} - l_{23}] [l_{31} - l_{43} - 1] \over
[2][l_{13} - l_{23}] [l_{23} + l_{43} + 3]}
\right)^{1/2}\\[4mm]
& &~~~~\times
(m;~m_{33}=-m_{23})_{(2)}^{-13-23+33+43+31},\\[6mm]
& &E_{32}(m;~m_{33}=-m_{23})_{(10)}^{-23+43}=
\\[4mm] & &~~~~
-\left({[l_{13} - l_{11}] [l_{23} + l_{31} + 3] \over
[l_{13} - l_{23}] [l_{23} + l_{43} + 4]}
\right)^{1/2}
(m;~m_{33}=-m_{23})_{(20)}^{-13-23+43+43+31}\\[4mm]
& &~~~~+ {1 \over [l_{23} + l_{43} + 4]}
\left({[l_{13} - l_{11}] [l_{31} - l_{43} - 1]
[l_{23} + l_{43} + 3] \over
[2][l_{13} - l_{23}]}
\right)^{1/2}\\[4mm]
& &~~~~\times
(m;~m_{33}=-m_{23})_{(2)}^{-13-23+33+43+31},\\[6mm]
& &E_{32}(m;~m_{33}=-m_{23})_{(20)}^{-13-23+43+43}=
\\[4mm] & &~~~~
\left({[l_{13} - l_{11} - 1] [l_{31} - l_{43} - 2] \over
[l_{13} - l_{23}] [l_{23} + l_{43} + 4]}
\right)^{1/2}
(m;~m_{33}=-m_{23})_{(21)}^{-13-13-23+33+43+43+31},\\[6mm]
& &E_{32}(m;~m_{33}=-m_{23})_{(2)}^{-13-23+33+43}=
\\[4mm] & &~~~~
\left({[2][l_{13} - l_{11} - 1] [l_{23} + l_{31} + 2] \over
[l_{13} - l_{23}] [l_{23} + l_{43} + 3]}
\right)^{1/2}
(m;~m_{33}=-m_{23})_{(21)}^{-13-13-23+33+43+43+31},\\[6mm]
& &E_{32}(m;~m_{33}=-m_{23})_{(11)}^{-13-13+33+43}=
\\[4mm] & &~~~~
-\left({[l_{11} - l_{23}] [l_{23} + l_{31} + 2] \over
[l_{13} - l_{23} - 1] [l_{23} + l_{43} + 3]}
\right)^{1/2}
(m;~m_{33}=-m_{23})_{(21)}^{-13-13-23+33+43+43+31},\\[6mm]
& &E_{32}(m;~m_{43}
=-m_{13})_{(21)}^{-13-13-23+33+43+43} = 0.\\
& &~~~~~~~~ ~~~ ~~~ ~~~ ~~ ~~~ ~~~ ~~ ~
{}~~~ ~~ ~~~ ~~~ ~~~ ~~~~ ~~~ ~~ ~~ ~~ ~~ ~~ ~~ ~~ ~~
{}~~ ~~~~ ~~ ~~ ~~ ~~ ~ ~~ ~~ ~~ ~~ ~~ ~~ ~~ ~~ ~~ (A.1)
\end{eqnarray*}

    The matrix elements of of $E_{23}$:

\begin{eqnarray*}
& &E_{23}(m;~m_{33}=-m_{23})_{(21)}^{-13-13-23+33+43+43}
=\\[4mm] & &~~~~
-\left({[l_{11} - l_{23}] [l_{23} + l_{31} + 1]
[l_{23} + l_{43} + 3] \over
[l_{13} - l_{23} - 1]}
\right)^{1/2}
(m;~m_{33}=-m_{23})_{(11)}^{-13-13+33+43-31}\\[4mm]
& &~~~~
+{\left([l_{13} - l_{11} - 1] [l_{31} - l_{43} - 3]
[l_{13} - l_{23}]\right)^{1/2} \over
[l_{23} + l_{43} + 4]}
(m;~m_{33}=-m_{23})_{(20)}^{-13-23+43+43-31}\\[4mm]
& &~~~~
+[l_{13} + l_{43} + 3]
{\left([2][l_{13} - l_{23}] [l_{13} - l_{11} - 1]
[l_{31} + l_{23} + 1] [l_{23} + l_{43} + 3]
\right)^{1/2} \over
[2][l_{13} - l_{23} - 1][l_{23} + l_{43} + 4]}
\\[4mm]
& &~~~~\times
(m;~m_{33}=-m_{23})_{(2)}^{-13-23+33+43-31}
,\\[6mm]
& &E_{23}(m;~m_{33}=-m_{23})_{(2)}^{-13-23+33+43}
=\\[4mm] & &~~~~
-[l_{23} + l_{43} + 4]
\left({[2][l_{11} - l_{23}] [l_{23} + l_{31} + 1]
 \over
[l_{13} - l_{23}][l_{23} + l_{43} + 3]}
\right)^{1/2}
(m;~m_{33}=-m_{23})_{(10)}^{-13+33-31}\\[4mm]
& &~~~~
+[l_{13} - l_{23} - 1]
\left({[2][l_{13} - l_{11}] [l_{31} - l_{43} - 2]
 \over
 [l_{13} - l_{23}] [l_{23} + l_{43} + 3]}
\right)^{1/2}
(m;~m_{33}=-m_{23})_{(10)}^{-23+43-31}\\[4mm]
& &~~~~
+
\left({[2][l_{11} - l_{23}] [l_{31} - l_{43} - 2]
 \over
 [l_{13} - l_{23}] [l_{23} + l_{43} + 3]}
\right)^{1/2}
(m;~m_{33}=-m_{23})_{(10)}^{-13+43-31}
,\\[6mm]
& &E_{23}(m;~m_{33}=-m_{23})_{(11)}^{-13-13+33+43}
=\\[4mm] & &~~~~
-[l_{13} + l_{43} + 3]
\left({[l_{13} - l_{11} - 1] [l_{23} + l_{31} + 1]
\over [l_{13} - l_{23} - 1] [l_{23} + l_{43} + 3]}
\right)^{1/2}
(m;~m_{33}=-m_{23})_{(10)}^{-13+33-31}\\[4mm]
& &~~~~
+[l_{13} - l_{23}]
\left({[l_{13} - l_{11} - 1] [l_{31} - l_{43} - 2]
 \over [l_{13} - l_{23} - 1[l_{23} + l_{43} +3]}
\right)^{1/2}
(m;~m_{33}=-m_{23})_{(10)}^{-13+43-31}
,\\[6mm]
& &E_{23}(m;~m_{33}=-m_{23})_{(20)}^{-13-23+43+43}
=\\[4mm] & &~~~~
-[l_{13} + l_{43} + 3]
\left({[l_{13} - l_{11}] [l_{31} + l_{23} + 2]
\over [l_{13} - l_{23}] [l_{23} + l_{43} + 4]}
\right)^{1/2}
(m;~m_{33}=-m_{23})_{(10)}^{-23+43-31}\\[4mm]
& &~~~~
+[l_{23} + l_{43} + 3]
\left({[l_{11} - l_{23}] [l_{23} + l_{31} + 2]
 \over [l_{13} - l_{23}] [l_{23} + l_{43} + 4]}
\right)^{1/2}
(m;~m_{33}=-m_{23})_{(10)}^{-13+43-31}
,\\[6mm]
& &E_{23}(m;~m_{33}=-m_{23})_{(10)}^{-23+43}
=\\[4mm] & &~~~~
- \left({[l_{11} - l_{23}] [l_{23} + l_{31} + 2]
[l_{23} + l_{43} + 3]
\over [l_{13} - l_{23}]}
\right)^{1/2}
(m;~m_{33}=-m_{23})_{(00)}^{-31}
,\\[6mm]
& &E_{23}(m;~m_{33}=-m_{23})_{(10)}^{-13+43}
=\\[4mm] & &~~~~
-[l_{13} + l_{43} + 3]
\left({[l_{13} - l_{11}] [l_{23} + l_{31} + 2]
\over [l_{13} - l_{23}] [l_{23} + l_{43} + 3]}
\right)^{1/2}
(m;~m_{33}=-m_{23})_{(00)}^{-31}
,\\[6mm]
& &E_{23}(m;~m_{33}=-m_{23})_{(10)}^{-13+33}
=\\[4mm] & &~~~~
- \left({[l_{13} - l_{11}] [l_{13} - l_{23}]
[l_{31} - l_{43} - 1]
\over [l_{23} + l_{43} + 3]}
\right)^{1/2}
(m;~m_{33}=-m_{23})_{(00)}^{-31}
,\\[6mm]
& &E_{23}(m;~m_{33}=-m_{23})_{(00)}=0.\\
& &~~~~~~~~ ~~~ ~~~ ~~~ ~~ ~~~ ~~~ ~~ ~
{}~~~ ~~ ~~~ ~~~ ~~~ ~~~~ ~~~ ~~ ~~ ~~ ~~ ~~
{}~~ ~~ ~~ ~~ ~~~~ ~~ ~~ ~~ ~~
{}~ ~~ ~~ ~~ ~~ ~~ ~~ ~~ (A.2)\\[2mm]
\end{eqnarray*}

    {\bf A.3. Class 3}\\

      The  matrix elements of $E_{32}$:

\begin{eqnarray*}
& &E_{32}(m;~m_{43}=-m_{23}+1)_{(00)}=\\[4mm] & &~~~~
-\left({[l_{13} - l_{11}] [l_{23} + l_{31} +3] \over
[l_{13} - l_{23}] [l_{23} + l_{33} + 3]} \right)^{1/2}
(m;~m_{43}=-m_{23}+1)_{(10)}^{-13+33+31}\\[4mm]
& &~~~~-\left({[l_{13} - l_{11}] [l_{33} - l_{31}] \over
[l_{13} - l_{23}] [l_{23} + l_{33} + 3]} \right)^{1/2}
(m;~m_{43}=-m_{23}+1)_{(10)}^{-13+43+31}\\[4mm]
& &~~~~-\left({[l_{11} - l_{23}] [l_{23} + l_{31}+3] \over
[l_{13} - l_{23}] [l_{23} + l_{33} + 3]} \right)^{1/2}
(m;~m_{43}=-m_{23}+1)_{(10)}^{-23++33+31},\\[6mm]
& &E_{32}(m;~m_{43}=-m_{23}+1)_{(10)}^{-13+33}=
\\[4mm] & &~~~~
\left({[l_{11} - l_{23}] [l_{23} + l_{31} + 3] \over
[l_{13} - l_{23}] [l_{23} + l_{33} + 4]}
\right)^{1/2}
(m;~m_{43}=-m_{23}+1)_{(20)}^{-13-23+33+33+31}\\[4mm]
& &~~~~-\left({[l_{13} - l_{11} - 1] [l_{33} - l_{31} + 1]
\over
[l_{13} - l_{23} - 1] [l_{23} + l_{33} + 3]} \right)^{1/2}
(m;~m_{43}=-m_{23}+1)_{(11)}^{-13-13+33+43+31}\\[4mm]
& &~~~~-{[l_{13} + l_{33} + 3] \over
[l_{13} - l_{23}-1] [l_{23} + l_{33} + 4]}
\left({[l_{11} - l_{23}]
[l_{33} - l_{31} + 1] \over
[2][l_{13} - l_{23}] [l_{23} + l_{33} + 3]}
\right)^{1/2}\\[4mm]
& &~~~~\times
(m;~m_{43}=-m_{23}+1)_{(3)}^{-13-23+33+43+31},\\[6mm]
& &E_{32}(m;~m_{43}=-m_{23}+1)_{(10)}^{-13+43}=
\\[4mm] & &~~~~
\left({[l_{13} - l_{11} - 1] [l_{23} + l_{31} + 2] \over
[l_{13} - l_{23} - 1] [l_{23} + l_{33} + 3]}
\right)^{1/2}
(m;~m_{43}=-m_{23}+1)_{(11)}^{-13-13+33+43+31}\\[4mm]
& &~~~~
+{1 \over [l_{13} - l_{23} - 1]}
\left({[l_{13} - l_{23}] [l_{11} - l_{23}]
[l_{23} + l_{31} + 2] \over
[2][l_{23} + l_{33} + 3]}
\right)^{1/2}\\[4mm]
& &~~~~
\times
(m;~m_{43}=-m_{23}+1)_{(3)}^{-13-23+33+43+31},\\[6mm]
& &E_{32}(m;~m_{43}=-m_{23}+1)_{(10)}^{-23+33}=
\\[4mm] & &~~~~
-\left({[l_{13} - l_{11}] [l_{23} + l_{31} + 3] \over
[l_{13} - l_{23}] [l_{23} + l_{33} + 4]}
\right)^{1/2}
(m;~m_{43}=-m_{23}+1)_{(20)}^{-13-23+33+33+31}\\[4mm]
& &~~~~+ {1 \over [l_{23} + l_{33} + 4]}
\left({[l_{13} - l_{11}] [l_{33} - l_{31} + 1]
[l_{23} + l_{33} + 3] \over
[2][l_{13} - l_{23}]}
\right)^{1/2}\\[4mm]
& &~~~~\times
(m;~m_{43}=-m_{23}+1)_{(3)}^{-13-23+33+43+31},\\[6mm]
& &E_{32}(m;~m_{43}=-m_{23}+1)_{(20)}{-13-23+33+33}=
\\[4mm] & &~~~~
-\left({[l_{13} - l_{11}-1] [l_{33} - l_{31} + 2] \over
[l_{13} - l_{23}] [l_{23} + l_{33} + 4]}
\right)^{1/2}
(m;~m_{43}=-m_{23}+1)_{(21)}^{-13-13-23+33+33+43+31},\\[6mm]
& &E_{32}(m;~m_{43}=-m_{23}+1)_{(3)}^{-13-23+33+43}=
\\[4mm] & &~~~~
\left({[2][l_{13} - l_{11} - 1] [l_{23} + l_{31} + 2] \over
[l_{13} - l_{23}] [l_{23} + l_{33} + 3]}
\right)^{1/2}
(m;~m_{43}=-m_{23}+1)_{(21)}^{-13-13-23+33+33+43+31},\\[6mm]
& &E_{32}(m;~m_{43}=-m_{23}+1)_{(11)}^{-13-13+33+43}=
\\[4mm] & &~~~~
-\left({[l_{11} - l_{23}] [l_{23} + l_{31} + 2] \over
[l_{13} - l_{23} - 1] [l_{23} + l_{33} + 3]}
\right)^{1/2}
(m;~m_{43}=-m_{23}+1)_{(21)}^{-13-13-23+33+33+43+31},\\[6mm]
& &E_{32}(m;~m_{43}
=-m_{13})_{(21)}^{-13-13-23+33+33+43} = 0.\\
& &~~~~~~~~ ~~~ ~~~ ~~~ ~~ ~~~ ~~~ ~~ ~
{}~~~ ~~ ~~~ ~~~ ~~~ ~~~~ ~~~ ~~ ~~ ~~ ~~ ~~ ~~ ~~ ~~
{}~~ ~~~~ ~~ ~~ ~~ ~~ ~ ~~ ~~ ~~ ~~ ~~ ~~ ~~ ~~ ~~ (A.3)
\end{eqnarray*}

    The matrix elements of of $E_{23}$:

\begin{eqnarray*}
& &E_{23}(m;~m_{43}=-m_{23}+1)_{(21)}^{-13-13-23+33+33+43}
=\\[4mm] & &~~~~
-\left({[l_{11} - l_{23}] [l_{23} + l_{31} + 1]
[l_{23} + l_{33} + 3] \over
[l_{13} - l_{23} - 1]}
\right)^{1/2}
(m;~m_{43}=-m_{23}+1)_{(11)}^{-13-13+33+43-31}\\[4mm]
& &~~~~
-\left({[l_{13} - l_{11} - 1] [l_{33} - l_{31} + 3]
[l_{13} - l_{23}] \over
[l_{23} + l_{33} + 4]}
\right)^{1/2}
(m;~m_{43}=-m_{23}+1)_{(20)}^{-13-23+33+33-31}\\[4mm]
& &~~~~
+[l_{13} + l_{33} + 3]
{\left([2][l_{13} - l_{11} - 1] [l_{13} - l_{23}]
[l_{23} + l_{31} + 1] [l_{23} + l_{33} + 3]
\right)^{1/2} \over
[2][l_{13} - l_{23} - 1][l_{23} + l_{33} + 4]}
\\[4mm]
& &~~~~\times
(m;~m_{43}=-m_{23}+1)_{(3)}^{-13-23+33+43-31}
,\\[6mm]
& &E_{23}(m;~m_{43}=-m_{23}+1)_{(3)}^{-13-23+33+43}
=\\[4mm] & &~~~~
-[l_{13} - l_{23} - 1]
\left({[2][l_{13} - l_{11}] [l_{33} - l_{31} + 2]
 \over
[l_{13} - l_{23}][l_{23} + l_{33} + 3]}
\right)^{1/2}
(m;~m_{43}=-m_{23}+1)_{(10)}^{-23+33-31}\\[4mm]
& &~~~~
-
\left({[2][l_{11} - l_{23}] [l_{33} - l_{31} + 2]
 \over
 [l_{13} - l_{23}] [l_{23} + l_{33} + 3]}
\right)^{1/2}
(m;~m_{43}=-m_{23}+1)_{(10)}^{-13+33-31}\\[4mm]
& &~~~~
+ [l_{23} + l_{33} + 4]
\left({[2][l_{11} - l_{23}] [l_{23} + l_{31} + 1]
 \over
 [l_{13} - l_{23}] [l_{23} + l_{33} + 3]}
\right)^{1/2}
(m;~m_{43}=-m_{23}+1)_{(10)}^{-13+43-31}
,\\[6mm]
& &E_{23}(m;~m_{43}=-m_{23}+1)_{(11)}^{-13-13+33+43}
=\\[4mm] & &~~~~
-[l_{13} - l_{23}]
\left({[l_{13} - l_{11} - 1] [l_{33} - l_{31} + 2]
\over [l_{13} - l_{23} - 1] [l_{23} + l_{33} + 3]}
\right)^{1/2}
(m;~m_{43}=-m_{23}+1)_{(10)}^{-13+33-31}\\[4mm]
& &~~~~
+[l_{13} + l_{33} + 3]
\left({[l_{13} - l_{11} - 1] [l_{23} + l_{31} + 1]
\over [l_{13} - l_{23} - 1[l_{23} + l_{33} + 3]}
\right)^{1/2}
(m;~m_{43}=-m_{23}+1)_{(10)}^{-13+43-31}
,\\[6mm]
& &E_{23}(m;~m_{43}=-m_{23}+1)_{(20)}^{-13-23+33+33}
=\\[4mm] & &~~~~
-[l_{13} + l_{33} + 3]
\left({[l_{13} - l_{11}] [l_{23} + l_{31} + 2]
\over [l_{13} - l_{23}] [l_{23} + l_{33} + 4]}
\right)^{1/2}
(m;~m_{43}=-m_{23}+1)_{(10)}^{-23+33-31}\\[4mm]
& &~~~~
+[l_{23} + l_{33} + 3]
\left({[l_{11} - l_{23}] [l_{23} + l_{31} + 2]
 \over [l_{13} - l_{23}] [l_{23} + l_{33}+4]}
\right)^{1/2}
(m;~m_{43}=-m_{23}+1)_{(10)}^{-13+33-31}
,\\[6mm]
& &E_{23}(m;~m_{43}=-m_{23}+1)_{(10)}^{-23+33}
=\\[4mm] & &~~~~
-\left({[l_{11} - l_{23}] [l_{23} + l_{31} + 2]
[l_{23} + l_{33} + 3]
\over [l_{13} - l_{23}]}
\right)^{1/2}
(m;~m_{43}=-m_{23}+1)_{(00)}^{-31}
,\\[6mm]
& &E_{23}(m;~m_{43}=-m_{23}+1)_{(10)}^{-13+43}
=\\[4mm] & &~~~~
-
\left({[l_{13} - l_{11}] [l_{13} - l_{23}]
[l_{33} - l_{31} + 1]
\over [l_{23} + l_{33} + 3]}
\right)^{1/2}
(m;~m_{43}=-m_{23}+1)_{(00)}^{-31}
,\\[6mm]
& &E_{23}(m;~m_{43}=-m_{23}+1)_{(10)}^{-13+33}
=\\[4mm] & &~~~~
-[l_{13} + l_{33} + 3]
\left({[l_{13} - l_{11}]
[l_{23} + l_{31} + 2]
\over [l_{13} - l_{23}] [l_{23} + l_{33} + 3]}
\right)^{1/2}
(m;~m_{43}=-m_{23}+1)_{(00)}^{-31}
,\\[6mm]
& &E_{23}(m;~m_{43}=-m_{23}+1)_{(00)}=0.\\
& &~~~~~~~~ ~~~ ~~~ ~~~ ~~ ~~~ ~~~ ~~ ~
{}~~~ ~~ ~~~ ~~~ ~~~ ~~~~ ~~~ ~~ ~~ ~~ ~~ ~~
{}~~ ~~ ~~ ~~ ~~~~ ~~ ~~ ~~ ~~
{}~ ~~ ~~ ~~ ~~ ~~ ~~ ~~ (A.4)\\[2mm]
\end{eqnarray*}

  {\bf A.4. Class 4}\\

      The  matrix elements of $E_{32}$:

\begin{eqnarray*}
& &E_{32}(m;~m_{33}=-m_{13}-1)_{(00)}=\\[4mm]
& &~~~~
-\left({[l_{13} - l_{11}] [l_{13} + l_{31} + 3] \over
[l_{13} - l_{23}] [l_{13} + l_{43} + 3]} \right)^{1/2}
(m;~m_{33}=-m_{13}-1)_{(10)}^{-13+43+31}\\[4mm]
& &~~~~-\left({[l_{11} - l_{23}] [l_{31} - l_{43}] \over
[l_{13} - l_{23}] [l_{13} + l_{43} + 3]} \right)^{1/2}
(m;~m_{33}=-m_{13}-1)_{(10)}^{-13+33+31}\\[4mm]
& &~~~~-\left({[l_{11} - l_{23}] [l_{13} + l_{31} + 3] \over
[l_{13} - l_{23}] [l_{13} + l_{43} + 3]} \right)^{1/2}
(m;~m_{33}=-m_{13}-1)_{(10)}^{-23+43+31},\\[6mm]
& &E_{32}(m;~m_{33}=-m_{13}-1)_{(10)}^{-13+43}=
\\[4mm] & &~~~~
\left({[l_{11} - l_{23}] [l_{13} + l_{31} + 3] \over
[l_{13} - l_{23}] [l_{13} + l_{43} + 4]}
\right)^{1/2}
(m;~m_{33}=-m_{13}-1)_{(20)}^{-13-23+43+43+31}\\[4mm]
& &~~~~-{1 \over (l_{13} + l_{43} + 4)}
\left({[l_{11} - l_{23}]
[l_{31} - l_{43} - 1]
[l_{13} + l_{43} + 3] \over
[2][l_{13} - l_{23}]}
\right)^{1/2}\\[4mm]
& &~~~~\times
(m;~m_{33}=-m_{13}-1)_{(4)}^{-13-23+33+43+31},\\[6mm]
& &E_{32}(m;~m_{33}=-m_{13}-1)_{(10)}^{-23+33}=
\\[4mm] & &~~~~
-\left({[l_{11} - l_{23} + 1] [l_{13} + l_{31} + 2] \over
[l_{13} - l_{23} + 1] [l_{13} + l_{43} + 3]}
\right)^{1/2}
(m;~m_{33}=-m_{13}-1)_{(11)}^{-23-23+33+43+31}\\[4mm]
& &~~~~+{1 \over [l_{13} - l_{23} + 1]}
\left({[l_{13} - l_{11}] [l_{13} - l_{23}]
[l_{13} + l_{31} + 2] \over
[2][l_{13} + l_{43} + 3]}
\right)^{1/2}\\[4mm]
& &~~~~
\times
(m;~m_{33}=-m_{13}-1)_{(4)}^{-13-23+33+43+31},\\[6mm]
& &E_{32}(m;~m_{33}=-m_{13}-1)_{(10)}^{-23+43}=
\\[4mm] & &~~~~
-\left({[l_{13} - l_{11}] [l_{13} + l_{31} + 3] \over
[l_{13} - l_{23}] [l_{13} + l_{43} + 4]}
\right)^{1/2}
(m;~m_{33}=-m_{13}-1)_{(20)}^{-13-23+43+43+31}\\[4mm]
& &~~~~
+ \left({[l_{11} - l_{23} - 1] [l_{31} - l_{43} - 1] \over
[l_{13} - l_{23} + 1] [l_{13} + l_{43} + 3]}
\right)^{1/2}
(m;~m_{33}=-m_{13}-1)_{(11)}^{-23-23+33+43+31}\\[4mm]
& &~~~~+ {[l_{23} + l_{43} + 3]
\over [l_{13} + l_{43} + 4] [l_{13} - l_{23} + 1]}
\left({[l_{13} - l_{11}] [l_{31} - l_{43} - 1]
\over
[2][l_{13} - l_{23}] [l_{13} + l_{43} + 3]}
\right)^{1/2}\\[4mm]
& &~~~~\times
(m;~m_{33}=-m_{13}-1)_{(4)}^{-13-23+33+43+31},\\[6mm]
& &E_{32}(m;~m_{33}=-m_{13}-1)_{(20)}^{-13-23+43+43}=
\\[4mm] & &~~~~
\left({[l_{11} - l_{23} + 1] [l_{31} - l_{43} - 2] \over
[l_{13} - l_{23}] [l_{13} + l_{43} + 4]}
\right)^{1/2}
(m;~m_{33}=-m_{13}-1)_{(21)}^{-13-23-23+33+43+43+31},\\[6mm]
& &E_{32}(m;~m_{33}=-m_{13}-1)_{(4)}^{-13-23+33+43}=
\\[4mm] & &~~~~
\left({[2][l_{11} - l_{23} + 1] [l_{13} + l_{31} + 2]
\over
[l_{13} - l_{23}] [l_{13} + l_{43} + 3]}
\right)^{1/2}
(m;~m_{33}=-m_{13}-1)_{(21)}^{-13-23-23+33+43+43+31},\\[6mm]
& &E_{32}(m;~m_{33}=-m_{13}-1)_{(11)}^{-23-23+33+43}=
\\[4mm] & &~~~~
\left({[l_{13} - l_{11}] [l_{13} + l_{31} + 2] \over
[l_{13} - l_{23} + 1] [l_{13} + l_{43} + 3]}
\right)^{1/2}
(m;~m_{33}=-m_{13}-1)_{(21)}^{-13-23-23+33+43+43+31},\\[6mm]
& &E_{32}(m;~m_{43}
=-m_{13})_{(21)}^{-13-23-23+33+43+43} = 0.\\
& &~~~~~~~~ ~~~ ~~~ ~~~ ~~ ~~~ ~~~ ~~ ~
{}~~~ ~~ ~~~ ~~~ ~~~ ~~~~ ~~~ ~~ ~~ ~~ ~~ ~~ ~~ ~~ ~~
{}~~ ~~~~ ~~ ~~ ~~ ~~ ~ ~~ ~~ ~~ ~~ ~~ ~~ ~~ ~~ ~~ (A.5)
\end{eqnarray*}

    The matrix elements of of $E_{23}$:

\begin{eqnarray*}
& &E_{23}(m;~m_{33}=-m_{13}-1)_{(21)}^{-13-23-23+33+43+43}
=\\[4mm] & &~~~~
\left({[l_{13} - l_{11}] [l_{13} + l_{31} + 1]
[l_{13} + l_{43} + 3] \over
[l_{13} - l_{23} - 1]}
\right)^{1/2}
(m;~m_{33}=-m_{13}-1)_{(11)}^{-23-23+33+43-31}\\[4mm]
& &~~~~
- \left({[l_{11} - l_{23} + 1] [l_{31} - l_{43} - 3]
[l_{13} - l_{23}] \over
[l_{13} + l_{43} + 4]}
\right)^{1/2}
(m;~m_{33}=-m_{13}-1)_{(20)}^{-13-23+43+43-31}\\[4mm]
& &~~~~
+{[l_{23} + l_{43} + 3]
\left([2][l_{11} - l_{23} + 1] [l_{13} - l_{23}]
[l_{13} + l_{31} + 1] [l_{13} + l_{43} + 3]
\right)^{1/2}
\over
[2][l_{13} - l_{23} + 1][l_{13} + l_{43} + 4]}
\\[4mm]
& &~~~~\times
(m;~m_{33}=-m_{13}-1)_{(4)}^{-13-23+33+43-31}
,\\[6mm]
& &E_{23}(m;~m_{33}=-m_{13}-1)_{(4)}^{-13-23+33+43}
=\\[4mm] & &~~~~
[l_{13} + l_{43} + 4]
\left({[2][l_{13} - l_{11}] [l_{13} + l_{31} + 1]
 \over
[l_{13} - l_{23}][l_{13} + l_{43} + 3]}
\right)^{1/2}
(m;~m_{33}=-m_{13}-1)_{(10)}^{-23+33-31}\\[4mm]
& &~~~~
-
\left({[2][l_{13} - l_{11}] [l_{31} - l_{43} - 2]
 \over
 [l_{13} - l_{23}] [l_{13} + l_{43} + 3]}
\right)^{1/2}
(m;~m_{33}=-m_{13}-1)_{(10)}^{-23+43-31}\\[4mm]
& &~~~~
+ [l_{13} - l_{23} + 1]
\left({[2][l_{11} - l_{23}] [l_{31} - l_{43} - 2]
 \over
 [l_{13} - l_{23}] [l_{13} + l_{43} + 3]}
\right)^{1/2}
(m;~m_{33}=-m_{13}-1)_{(10)}^{-13+43-31}
,\\[6mm]
& &E_{23}(m;~m_{33}=-m_{13}-1)_{(11)}^{-23-23+33+43}
=\\[4mm] & &~~~~
-[l_{23} + l_{43} + 3]
\left({[l_{11} - l_{23} + 1] [l_{13} + l_{31} + 1]
\over [l_{13} - l_{23} + 1] [l_{13} + l_{43} + 3]}
\right)^{1/2}
(m;~m_{33}=-m_{13}-1)_{(10)}^{-23+33-31}\\[4mm]
& &~~~~
+ [l_{13} - l_{23}]
\left({[l_{11} - l_{23} + 1] [l_{31} - l_{43} - 2]
\over [l_{13} - l_{23} + 1[l_{13} + l_{43} +3]}
\right)^{1/2}
(m;~m_{33}=-m_{13}-1)_{(10)}^{-23+43-31}
,\\[6mm]
& &E_{23}(m;~m_{33}=-m_{13}-1)_{(20)}^{-13-23+43+43}
=\\[4mm] & &~~~~
-[l_{13} + l_{43} + 3]
\left({[l_{13} - l_{11}] [l_{13} + l_{31} + 2]
\over [l_{13} - l_{23}] [l_{13} + l_{43} + 4]}
\right)^{1/2}
(m;~m_{33}=-m_{13}-1)_{(10)}^{-23+43-31}\\[4mm]
& &~~~~
+[l_{23} + l_{43} + 3]
\left({[l_{11} - l_{23}] [l_{13} + l_{31} + 2]
 \over [l_{13} - l_{23}] [l_{13} + l_{43} + 4]}
\right)^{1/2}
(m;~m_{33}=-m_{13}-1)_{(10)}^{-13+43-31}
,\\[6mm]
& &E_{23}(m;~m_{33}=-m_{13}-1)_{(10)}^{-23+43}
=\\[4mm] & &~~~~
- [l_{23} + l_{43} + 3]
\left({[l_{11} - l_{23}] [l_{13} + l_{31} +2]
\over [l_{13} - l_{23}] [l_{13} + l_{43} + 3]}
\right)^{1/2}
(m;~m_{33}=-m_{13}-1)_{(00)}^{-31}
,\\[6mm]
& &E_{23}(m;~m_{33}=-m_{13}-1)_{(10)}^{-23+33}
=\\[4mm] & &~~~~
\left({[l_{11} - l_{23}] [l_{13} - l_{23}]
[l_{31} - l_{43} - 1]
\over  [l_{13} + l_{43} + 3]}
\right)^{1/2}
(m;~m_{33}=-m_{13}-1)_{(00)}^{-31}
,\\[6mm]
& &E_{23}(m;~m_{33}=-m_{13}-1)_{(10)}^{-13+43}
=\\[4mm] & &~~~~
- \left({[l_{13} - l_{11}] [l_{13} + l_{31} + 2]
[l_{13} + l_{43} + 3]
\over [l_{13} - l_{23}]}
\right)^{1/2}
(m;~m_{33}=-m_{13}-1)_{(00)}^{-31}
,\\[6mm]
& &E_{23}(m;~m_{33}=-m_{13}-1)_{(00)}=.0\\
& &~~~~~~~~ ~~~ ~~~ ~~~ ~~ ~~~ ~~~ ~~ ~
{}~~~ ~~ ~~~ ~~~ ~~~ ~~~~ ~~~ ~~ ~~ ~~ ~~ ~~
{}~~ ~~ ~~ ~~ ~~~~ ~~ ~~ ~~ ~~
{}~ ~~ ~~ ~~ ~~ ~~ ~~ ~~ (A.6)\\[2mm]
\end{eqnarray*}

  {\bf A.5. Class 5}\\

      The  matrix elements of $E_{32}$:

\begin{eqnarray*}
& &E_{32}(m;~m_{33}=-m_{23},~m_{43}=-m_{13})_{(00)}=
\\[4mm] & &~~~~
- {
\left([l_{13} - l_{11}] [l_{13} + l_{31} + 3]
\right)^{1/2} \over [l_{13} - l_{23}]}
(m;~m_{33}=-m_{23},~m_{43}=-m_{13})_{(10)}^{-13+33+31}\\[4mm]
& &~~~~
- {
\left([l_{11} - l_{23}] [l_{23} + l_{31} + 3]
\right)^{1/2} \over [l_{13} - l_{23}]}
(m;~m_{33}=-m_{23},~m_{43}=-m_{13})_{(10)}^{-23+43+31},\\[6mm]
& &E_{32}(m;~m_{33}=-m_{23},~m_{43}=-m_{13})_{(10)}^{-13+33}=
\\[4mm] & &~~~~
-
{\left([2][l_{11} - l_{23}] [l_{23} + l_{31} + 2]
\right)^{1/2} \over [2][l_{13} - l_{23} - 1]}
(m;~m_{33}=-m_{23},~m_{43}=-m_{13})_{(5)}^{-13-23+33+43+31},\\[6mm]
& &E_{32}(m;~m_{33}=-m_{23},~m_{43}=-m_{13})_{(10)}^{-23+43}=
\\[4mm] & &~~~~
{\left([2][l_{13} - l_{11}] [l_{13} + l_{31} + 2]
\right)^{1/2} \over [2][l_{13} - l_{23} - 1]}
(m;~m_{33}=-m_{23},~m_{43}=-m_{13})_{(5)}^{-13-23+33+43+31},\\[6mm]
& &E_{32}(m;~m_{33}=
-m_{23},~m_{43}=-m_{13})_{(5)}^{-13-23+33+43}= 0.\\
& &~~~~~~~~ ~~~ ~~~ ~~~ ~~ ~~~ ~~~ ~~ ~
{}~~~ ~~ ~~~ ~~~ ~~~ ~~~~ ~~~ ~~ ~~ ~~ ~~ ~~ ~~ ~~ ~~
{}~~ ~~~~ ~~ ~~ ~~ ~~ ~ ~~ ~~ ~~ ~~ ~~ ~~ ~~ ~~ ~~ (A.7)
\end{eqnarray*}

    The matrix elements of of $E_{23}$:

\begin{eqnarray*}
& &E_{23}(m;~m_{33}=-m_{23},~m_{43}=-m_{13})_{(5)}^{-13-23+33+43}
=\\[4mm] & &~~~~
{[l_{13} - l_{23} - 1] \over [l_{13} - l_{23}]}
\left([2][l_{13} - l_{11}] [l_{13} + l_{31} + 1]
\right)^{1/2}\\[4mm]
& &~~~~
\times
(m;~m_{33}=-m_{23},~m_{43}=-m_{13})_{(10)}^{-23+43-31}\\[4mm]
& &~~~~
+{[l_{13} - l_{23} - 1] \over [l_{13} - l_{23}]}
\left([2][l_{11} - l_{23}] [l_{23} + l_{31} + 1]
\right)^{1/2}\\[4mm]
& &~~~~
\times
(m;~m_{33}=-m_{23},~m_{43}=-m_{13})_{(10)}^{-13+33-31},\\[6mm]
& &E_{23}(m;~m_{33}=-m_{23},~m_{43}=-m_{13})_{(10)}^{-23+43}
=\\[4mm] & &~~~~
\left([l_{11} - l_{23}] [l_{23} + l_{31} + 2]
\right)^{1/2}
(m;~m_{33}=-m_{23},~m_{43}=-m_{13})_{(00)}^{-31},\\[6mm]
& &E_{23}(m;~m_{33}=-m_{23},~m_{43}=-m_{13})_{(10)}^{-13+33}
=\\[4mm] & &~~~~
- \left([l_{13} - l_{11}] [l_{13} + l_{31} + 2]
\right)^{1/2}
(m;~m_{33}=-m_{23},~m_{43}=-m_{13})_{(00)}^{-31},\\[6mm]
& &E_{23}(m;~m_{33}=-m_{23},~m_{43}=-m_{13})_{(00)}=0.\\
& &~~~~~~~~ ~~~ ~~~ ~~~ ~~ ~~~ ~~~ ~~ ~
{}~~~ ~~ ~~~ ~~~ ~~~ ~~~~ ~~~ ~~ ~~ ~~ ~~ ~~
{}~~ ~~ ~~ ~~ ~~~~ ~~ ~~ ~~ ~~
{}~ ~~ ~~ ~~ ~~ ~~ ~~ ~~ (A.8)\\[2mm]
\end{eqnarray*}
\begin{flushleft}
{\bf REFERENCES}
\end{flushleft}
\vspace*{2mm}
\begin{enumerate}
\item Nguyen Anh Ky, J. Math. Phys. {\bf 35}, 2583 (1994).
\item P. P. Kulish and N. Yu. Reshetikhin,
{\it Lett. Math. Phys.} {\bf 18},
143 (1989) ;  E. Celeghini, Tch. Palev and
M.  Tarlini,  Mod.
Phys. Lett. {\bf B 5}, 187 (1991).
\item Tch. D. Palev and V. N. Tolstoy,
 Comm. Math. Phys.
{\bf 141}, 549 (1991).
\item R. Floreanini, V. Spiridonov and L. Vinet,
Comm. Math. Phys.
{\bf 137}, 149 (1991);
E. D'Hoker,  R. Floreanini and L. Vinet,
J. Math.  Phys., {\bf 32}, 1427 (1991);
Tch. Palev J.Math.Phys. {\bf 34}, 4872 (1993)..
\item R. B. Zhang,  J. Math.  Phys.,
{\bf 34}, 1236 (1993).
\item Tch. Palev, N. Stoilova and J. Van der Jeugt,
Preprint IC/93/157, to appear in Comm. Math. Phys..
\item V. D. Drinfel'd, "Quantum groups", in
{\it Proceedings of the International Congress of
Mathematicians}, 1986, Berkeley, vol. {\bf 1}, 798-820
(The American Mathematical Society, 1987).
\item M. Jimbo, Lett. Math. Phys. {\bf 10}, 63 (1985);
{\it ibit} {\bf 11}, 247 (1986).
\item L. D. Faddeev, N. Yu. Reshetikhin and L. A. Takhtajan,
Algebra and Analys, {\bf 1}, 178  (1987).
\item Yu. I. Manin, {\it Quantum groups and
non-commutative geometry}, Centre des Recherchers
Math\'ematiques, Montr\'eal (1988); {\it Topics
in non-commutative geometry}, Princeton University
Press, Princeton, New Jersey (1991).
\item S. I. Woronowicz, Comm. Math. Phys., {\bf 111},
613 (1987); {\bf 122}, 125 (1989).
\item R. Floreanini, D. Leites and L. Vinet,
Lett. Math. Phys.
{\bf 23}, 127 (1991);
M. Scheunert, Lett. Math. Phys. {\bf 24}, 173 (1992);
S. M. Khoroshkin and V. N. Tolstoy, Comm. Math. Phys.
{\bf 141}, 599 (1991).
\item V. Kac, Comm. Math. Phys., {\bf 53}, 31 (1977);
Adv. Math. {\bf 26}, 8 (1977);
{\it Lecture Notes in Mathematics}
{\bf 676}, 597 (Springer - Verlag, Berlin 1978).
\item A. H. Kamupingene, Nguyen Anh Ky and
Tch. D. Palev, J. Math. Phys. {\bf 30}, 553 (1989).
\item Tch. Palev and N. Stoilova,
J. Math.  Phys., {\bf 31}, 953 (1990).
\item I. M. Gel'fand and M. L. Zetlin,
{\it Dokl. Akad. Nauk USSR}, {\bf 71},
825 (1950), (in Russian); for a detailed
description of the Gel'fand-Zetlin basis see also
G. E. Baird and L. C. Biedenharn, {\it J. Math.  Phys.},
{\bf 4}, 1449 (1963);
A. O. Barut and R. Raczka,
{\it Theory of Group Representations and Applications},
Polish Scientific Publishers, Warszawa, 1980.
\item M. Rosso, Comm. Math. Phys. {\bf 117}, 581 (1987).
\item G. Lusztig, Adv. in Math {\bf 70}, 237 (1988).
\item H. Schlosser, Seminar Sophus Lie {\bf 3},15 (1993).
\item Tch. Palev, J. Math. Phys. {\bf 27}, 1994 (1986).
\item Tch. Palev,  J. Math. Phys. {\bf 28}, 2280 (1987);
Funct. Anal. Appl. {\bf 21}, 245 (1987).
\item Tch. Palev, J. Math. Phys. {\bf 29}, 2589 (1988);
J. Math. Phys. {\bf 30}, 1433 (1989).
\item Tch. Palev, Funct. Anal. Appl. {\bf 23}, 141 (1989).
\end{enumerate}
\end{document}